\documentclass[fleqn,usenatbib]{mnras}
\usepackage{adjustbox}
\usepackage[T1]{fontenc}
\usepackage{graphicx}   % Including figure files

\usepackage{savesym}
\usepackage{amsmath}
\savesymbol{iint}
\usepackage{txfonts}
\restoresymbol{TXF}{iint}

 \usepackage{siunitx}

\usepackage{hyperref}
\title[Mrk~110 continuum reverberation lags]{On the multi-wavelength variability of Mrk~110: Two components acting at different timescales}

\author[F. M. Vincentelli et. al]{F. M.  Vincentelli$^{1}$\thanks{E-mail:F.M.Vincentelli@soton.ac.uk}, I. McHardy $^{1}$, E. M. Cackett$^{2}$, A. J. Barth$^{3}$, K. Horne$^{4}$, 
\newauthor
 M. Goad$^{5}$, K. Korista$^{6}$, J. Gelbord$^{7}$, W. Brandt$^{8,9,10}$,  R. Edelson$^{11}$, J. A. Miller$^2$, 
 \newauthor
M. Pahari$^{12}$,  B.M. Peterson$^{12,13}$, T. Schmidt$^{14}$, R.~D. Baldi$^{15,1}$, E. Breedt$^{16}$, 
  \newauthor
 J.~V. Hern{\'a}ndez Santisteban$^{4}$,  E. Romero-Colmenero$^{17,18}$,  M. Ward$^{19}$, \newauthor D. R. A. Williams$^{20}$\\
$^{1}$Department of Physics and Astronomy, University of Southampton, SO17 1BJ, UK\\
$^{2}$Wayne State University, Department of Physics \& Astronomy, 666 W Hancock St, Detroit, MI 48201, USA\\
$^{3}$Department of Physics and Astronomy, 4129 Frederick Reines Hall, University of California, Irvine, CA, 92697-4575, USA\\
$^{4}$SUPA Physics and Astronomy, University of St. Andrews, North Haugh, KY16 9SS, UK\\
$^{5}$School of Physics and Astronomy, University of Leicester, Leicester, LE1 7RH, UK\\
$^{6}$Department of Physics, Western Michigan University, 1120 Everett Tower, Kalamazoo, MI 49008-5252, USA\\
$^{7}$Spectral Sciences Inc., 4 Fourth Avenue, Burlington, MA 01803, USA\\
$^8$Department of Astronomy and Astrophysics, The Pennsylvania State University, 525 Davey Lab, University Park, PA 16802, USA\\
$^9$Institute for Gravitation and the Cosmos, The Pennsylvania State University, University Park, PA 16802, USA\\
$^{10}$Department of Physics, 104 Davey Lab, The Pennsylvania State University, University Park, PA 16802, USA\\
$^{11}$Department of Astronomy, University of Maryland, College Park, MD 20742-2421, USA\\
$^{12}$Department of Physics, Indian Institute of Technology, Hyderabad 502285, India\\
$^{13}$Center for Cosmology and AstroParticle Physics; The Ohio State University; 192 West Woodruff Ave., Columbus, OH 43210, USA \\
$^{14}$ Department of Physics and Astronomy, University of California, Los Angeles, CA 90095-1547, USA\\
$^{15}$ INAF - Istituto di Radioastronomia, Via P. Gobetti 101, I-40129 Bologna, Italy\\
$^{16}$Institute of Astronomy, University of Cambridge, Madingley Road, Cambridge, CB3 0HA, UK\\
$^{17}$South African Astronomical Observatory, P.O Box 9, Observatory 7935, Cape Town, South Africa\\
$^{18}$Southern African Large Telescope Foundation, P.O Box 9, Observatory 7935, Cape Town, South Africa\\
$^{19}$Centre for Extragalactic Astronomy, Department of Physics, University of Durham, South Road, Durham, DH1 3LE, UK\\
$^{20}$Jodrell Bank Centre for Astrophysics, School of Physics and Astronomy,The University of Manchester, Manchester, M13 9PL, UK\\
}

\date{Accepted XXX. Received YYY; in original form ZZZ}

\pubyear{2021}
\begin{document}

\label{firstpage}
\pagerange{\pageref{firstpage}--\pageref{lastpage}}
\maketitle

% Abstract of the paper
\begin{abstract}
We present the first intensive continuum reverberation mapping study of the high accretion-rate Seyfert galaxy Mrk~110. The source was monitored almost daily for more than 200 days with
the \textit{Swift} X-ray and UV/optical telescopes, supported by 
ground-based observations from Las Cumbres Observatory, the Liverpool Telescope, and the Zowada Observatory, thus extending the
wavelength coverage to 9100\,\AA. 
 Mrk~110 was found to be significantly variable at all wavebands. Analysis of  the intraband lags   reveals  two different behaviours, depending on the timescale. On  timescales { shorter} than 10 days the lags, relative to 
  the shortest UV waveband ($\sim1928$\,\AA), increase with increasing wavelength up to a maximum of $\sim2$d lag for the
 longest waveband ($\sim9100$\,\AA), 
 consistent with the expectation from disc reverberation. On longer timescales, however, the g-band lags the Swift BAT hard X-rays by $\sim10$ days, with the z-band lagging the g-band by a similar amount, which cannot be explained in terms of simple reprocessing from the accretion disc. We interpret this result as an interplay between the emission from the accretion disc and diffuse continuum radiation from the broad line region.

\end{abstract}

% Select between one and six entries from the list of approved keywords.
% Don't make up new ones.
\begin{keywords}
accretion, accretion disc --- galaxies: Seyfert --- black hole physics --- X-rays: galaxies --- galaxies: individual: Mrk~110
\end{keywords}

%%%%%%%%%%%%%%%%%%%%%%%%%%%%%%%%%%%%%%%%%%%%%%%%%%

%%%%%%%%%%%%%%%%% BODY OF PAPER %%%%%%%%%%%%%%%%%%
 
\section{Introduction} 

%{\bf Introduction needs a bit of reworking. Standard plan is to have a short intro to what an AGN is, and briefly what we've learned recently (eg with \textit{Swift}) and then get rapidly onto the problems that we've discovered, like the X-ray/uv offset and the variations on different timescales. Try to bring in some big picture stuff about understanding the overall inner geometry of AGN. Then say what programmes we have just done which are designed to address these probs. Finish with v. brief outline of structure of the paper. Some of this stuff is already in the intro but it needs to be a bit more direct about the problems, ie proving motivation. You have the accretion-rate in there but we need some text to say why studying higher mdot is important. Maybe refer to the end of my 2018 paper where excess v-band to uv lags seems to get less in higher temp discs, or something like that.}

 Despite decades of observations across almost all of the 
 accessible  electromagnetic spectrum, and their impact in   galaxy  evolution \citep[e.g.,][]{ferrarese2000,gebhardt2000,marconi}, many aspects  of  active galactic nuclei (AGN) are still poorly understood. 
 Whilst it is clear that the origin of emission from AGN is fundamentally 
 the result of accretion of matter onto supermassive black holes
 (SMBHs;  $10^6-10^9 $M$_\odot$) at the centre of galaxies \citep{lyndenbel969,rees1984,eht2019_bh}, understanding their detailed inner
 geometry remains challenging  
 %due to the complex interplay between multiple physical components 
 \citep{padovani2017}. 
 
 Most of the AGN luminosity is released as thermal radiation from an optically thick, geometrically thin accretion disc which dominates the UV, and a non-thermal power-law which dominates the hard X-rays and which arises from the Compton up-scattering of lower energy seed photons \citep{shakura1973,haardt_maraschi1991}. It is commonly accepted that the non-thermal component originates from a geometrically thick and optically thin region, often called the corona, but its geometry remains unclear \citep[e.g.,][]{done2012,petrucci2018,arcodia2019}.  
 Determining the geometry of the accretion flow and inner corona is one of the main goals in the study of AGN. This geometry can be mapped out using the lags between different wavebands: a process known as reverberation mapping \citep{blandford1982,peterson1993}. Initially this technique was employed by measuring the lags between the continuum and the emission lines in the UV and optical bands, thereby measuring the size of the broad line region (BLR) and the mass of the SMBH \citep[ e.g.,][]{peterson2004,bentz2013}. However measurement of lags between X-ray bands can also reveal the mass and spin of the SMBH and also the size and geometry of the X-ray emitting region  \citep{demarco2013,cackett2014,Emmanoulopoulos2014,kara2016,caballerogarcia2018,ingram2019}.

The origin of the variability in the UV and optical bands and their relationship to the  X-ray emission has been a matter of considerable debate for several years. Do variations propagate inwards, e.g. as UV seed photon variations or accretion-rate variations \citep{arevalo2006} or outwards, by reprocessing X-rays by surrounding material? Measurement of the lag between the X-ray and UV/optical bands should decide. Monitoring campaigns based mainly on RXTE X-ray observations and ground based optical observations \citep[e.g. ][]{uttley2003,suganuma2006,arevalo2008,arevalo2009,breedt2009,breedt2010,lira2011,camerom2012} generally indicate that the optical lags the X-rays by about a day, consistent with reprocessing of X-rays by a surrounding accretion disc, but with too large uncertainty in any single AGN to be absolutely certain.  \citet{sergeev2005} showed that longer wavelength optical bands lagged behind the B-band with lags increasing with wavelength and \citet{cackett2007} showed that the lags within the optical bands were consistent with the predictions of reprocessing by an accretion disc with the temperature profile defined by \citet{shakura1973}: i.e. lag $\tau \propto \lambda^{4/3}$.

Intensive monitoring with Swift and other facilities has now greatly improved our general understanding of these sources \citep[e.g.,][]{shappee,mchardy2014,edelson2015,fausnaugh2016,edelson2017,cackett2018,mchardy2018,edelson2019}. Reprocessing of high-energy radiation by an accretion disc has been shown to be a significant contributor to UV and optical variability, but these observations have highlighted problems with the straightforward \citet{shakura1973} disc model. In particular, the observed lags are $\sim2-3 \times$ longer than expected \citep[e.g.,][]{mchardy2014}, implying a larger than expected disc, in agreement with previous observations of microlensing \citep{morgan2010}. The U vs UVW2 band lag are also larger than expected \citep[e.g.,][]{cackett2018}. In addition, almost all observations show that the X-rays lead the UV by considerably longer than would be expected purely on the basis of direct reprocessing by an accretion disc. Finally,  long-timescale lightcurves sometimes have shown  in the UV/optical trends which are not paralleled in the X-rays, implying an additional source of UV/optical variability \citep[e.g.][]{breedt2009,mchardy2014,hernandez,kammoun2021}.

A number of possible solutions to these problems have been proposed. The over-large discs might be explained if the discs are clumpy \citep{dexter2011}. The excessive lag in the U-band can
be explained as arising from Balmer continuum emission from the BLR \citep{korista2001,korista2019}. The excessive lag between the X-ray and UV bands might be explained if the X-rays do not directly illuminate the disc but first scatter slowly through the inflated inner edge of the disc, emerging as far-UV radiation which then illuminates the outer disc \citep{gardner2017}. Alternatively, contributions from the BLR can also explain the additional lag in at least one case \citep{mchardy2018}. There are additional, well-known, problems with reprocessing from a disc in that the observed UV/optical lightcurves are much \lq\lq smoother\rq\rq\ than expected. The model of \citet{gardner2017}, with an extended far-UV illuminating source, provides one solution, as does a very large  X-ray source \citep[e.g][]{arevalo2008,kammoun}, although the required source size ($\sim100R_{G}$) is much larger than measured ($\sim4-5R_{G}$) by X-ray reverberation methods \citep[e.g][]{Emmanoulopoulos2014, cackett2014} or X-ray eclipses \citep[][]{gallo2021}.

Interestingly, the large majority of AGN monitored to date have a broadly similar accretion-rate. This parameter is known to have a crucial role in the configuration of the accretion flow of all accreting compact objects \citep[e.g.,][]{ done2012,marcel2018,noda2018,koljonen}.  Analysis presented by  \citet{mchardy2018} indicated a possible dependence of the lag on the disc temperature, finding a smaller discrepancy between the observed and expected ratio of X-ray-to-UV and UV-to-optical lags in the systems with hotter discs, motivating further investigation at higher accretion-rates. However, the few studies to date on higher accretion-rate objects \citep{pahari2020,cackett2020} still reported discrepancies between the observed and predicted lags similar to those of lower accretion-rate AGN. 

Among the various classes of AGN, narrow-line Seyfert~1 (NLS1) galaxies
\citep[i.e.  active galaxies with optical emission lines width at half maximum (FWHM) $\approx82000$~km~s$^{-1}$, weak  $[$\ion{O}{III}$\rbrack$, and a strong \ion{Fe}{II}/\ion{H}{$\beta$} ratio; ][]{osterbrock,manthur}
and are therefore probably one of the most suited   for  this  investigation. Not only do they have a relatively low mass, and therefore a significant higher variability amplitude, but also they typically show a high accretion-rate  \citep{nicastro2000,veron-cetty}.  Here we present observations of another high accretion-rate AGN: Mrk~110. This source, despite the presence of strong  $[$\ion{O}{III}$\rbrack$ lines, and the very weak \ion{Fe}{II}, has been classified as a NSL1 due narrow Balmer lines \citep[FWHM$_\ion{H}{$\beta$}$=1800~km~s$^{-1}$; see also][ for a discussion]{kollatschny2001,veron-cetty110}. Different accretion rate measurements have been reported depending on the method \citep[see e.g.][]{Meyer-Hofmeister,dallabonta2020}. For the purposes of this paper we take the value of $L/L_{\rm Edd}\approx$40\%  \citep{Meyer-Hofmeister}. The exact value does not affect the conclusions of this paper.
It has a black hole mass $2\times10^7$ M$_\odot$  \citep{peterson1998,kollatschny2001,kollatschny2003,peterson2004,bentz2015} and is  150 Mpc ($z=0.03552$) distant. It is known to be one of the most variable AGN in the optical band, with variations of $\sim2 \times$ on a timescale of a few months \citep{peterson1998,bischoff1998}. The few optical line reverberation mapping studies on Mrk~110 achieved weekly sampling over a period of several months and revealed a relatively long lag ($\approx$25 days) for the  H$\beta$ line, as expected from a high luminosity source \citep{kaspi2000}, and indications for stratification depending on the ionization of the source \citep{kollatschny2001}. Here we present results from an intensive multiwavelength campaign of $\approx$200 days combining observations  from both space and ground-based telescopes.

%These strong discrepancies led rapidly to the development of alternative physical scenarios from the usual irradiated  accretion disc. In particular,   models invoked  to explain the observed lags   include  a stratified geometrically thick accretion disc \citep{gardner2017,edelson2017}, higher illuminating sources \citep{kammoun}, and  dominant contribution from the broad line region \citep{korista2001,cackett2018,mchardy2018,cheleouce2019,korista2019}. The presence of more than one valid physical scenario may be evidence of a possible evolution of the accretion flow geometry depending on some other physical parameter. 
%In particular, accretion-rate is known to have a crucial role for the configuration of the accretion flow of all accreting compact objects \citep[e.g.,][]{ done2012,marcel2018,noda2018,koljonen}. Therefore, motivated also by the limited range of accretion-rate of the sources monitored with the {\it Neil Gehrels Swift Observatory} (\textit{Swift} from hereafter), in the last few years attention started to focus especially on higher accretion-rate objects \citep{pahari2020,cackett2020}.  

%In particular, for this paper, we focused our attention on Mrk~110: 

\section{Observations and Light Curve Measurements}

\begin{figure*} 
\centering
\includegraphics[width=1.5\columnwidth]{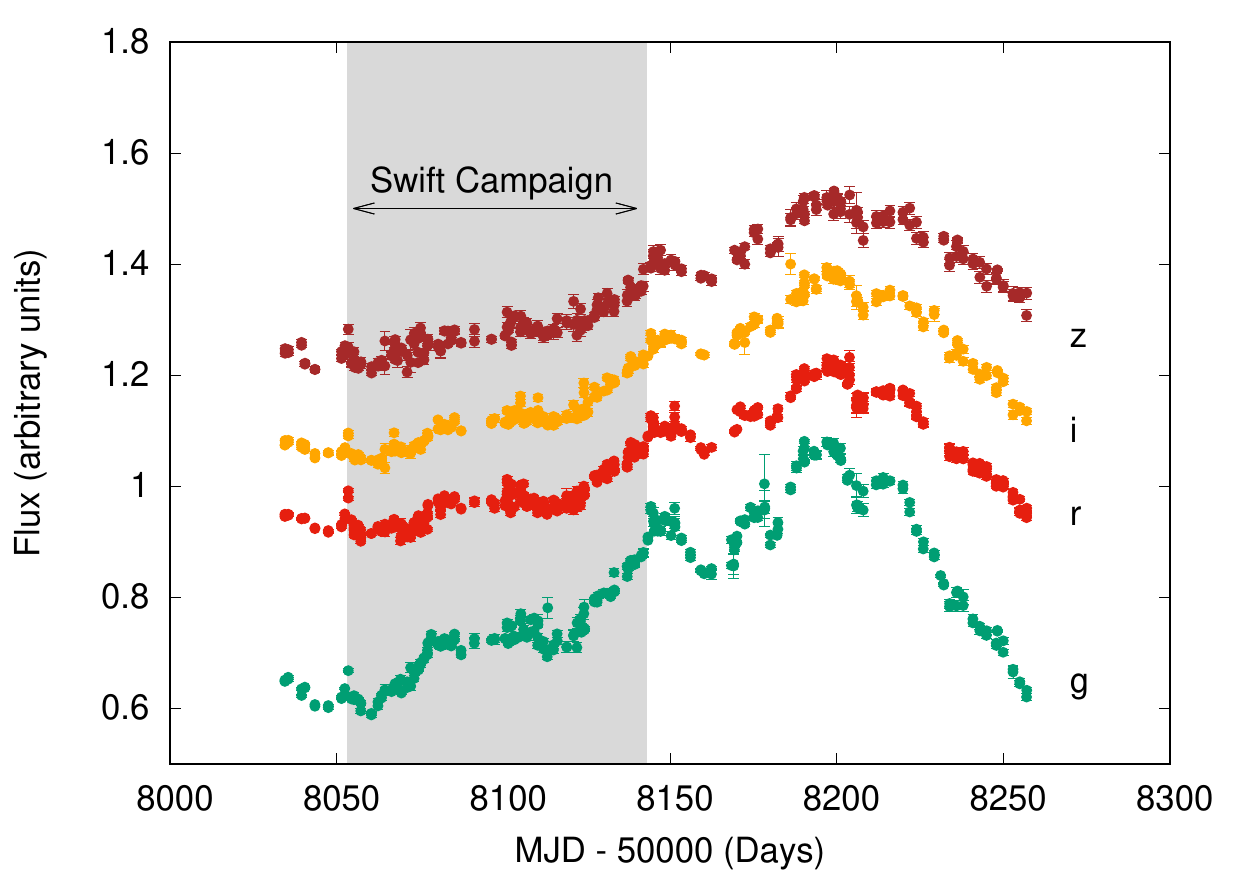}
\caption{LCO+LT observations of Mrk~110 using  \textit{griz} filters. Lightcurves were computed with  respect to their mean, and shifted vertically for clarity. The grey area shows the strictly simultaneous coverage with  \textit{Swift} observations.} 
\label{fig:lc_overall}
\end{figure*}

Mrk~110 was monitored by \textit{Swift} in X-rays and in 6 UV and optical bands for 3 months from 26 October 2017 to 25 January 2018. Optical imaging observations were also obtained over a longer monitoring duration from ground-based facilities including \textit{Las Cumbres Observatory} (LCO), the \textit{Liverpool Telescope} (LT), and the {\it Zowada Observatory}.  The log of these observations is given in Table~1. In Fig \ref{fig:lc_overall} we show the overall lightcurve of the source from LCO+LT, highlighting the strictly simultaneous window with the \textit{Swift} campaign (shown in Fig. \ref{fig:lc_short}). The \textit{Swift} and ground-based observations are described below (Sections~\ref{Swift} and ~\ref{groundbased}).
 
\begin{table*}

        \caption{Report of the observations. Columns 1 and 2 report telescope and filter respectively, 3 and 4  show the start and end of the lighcurve in MJD. Column 5 reports the number visits during the whole campaign.  The average, minimum and maximum flux  (in units of count rate for XRT, and ($\times 10^{-15}$) erg cm$^{-2}$ s$^{-1}$ \si{\angstrom}$^{-1}$ for the other filters)  during   the period of   \textit{strict simultaneity}  of all observatories  (i.e. MJD 58043-58153) are reported in column 6, 7 and 8. Column 9 reports the number of points during the  XRT campaign.}

\begin{tabular}{ l c | c c c | c c c  r }
& & & & & & (MJD 58053-58143) &\\

Telescope & Filter & Start & End & No. Visits & Avg. & Min. & Max. & No. Visits \\ 

\hline
XRT & 0.3-10 keV & 58053.1 & 58143.9 & 253 & 1.24 $\pm$ 0.23 & 0.66  & 1.89 & 253 \\ 
UVOT & UVW2& 58053.1 & 58143.9 &  226 & 23 $\pm$ 4 & 14.75 & 33  & 226\\
- & UVM2 & 58053.1 & 58143.9 &  224 & 20 $\pm $ 3 & 13.6 & 26.8 & 224\\

- & UVW1 & 58053.1 & 58143.9 &  233 & 17 $\pm $ 2 & 12.6 & 22.7 & 233\\
-  & U & 58053.1 & 58143.9 &  243 & 11.1  $\pm$ 1  &  8.6 & 13.7 & 243\\ 
- & B  & 58053.1 & 58143.9 &  244 & 6   $\pm$ 0.5 & 4.9 & 7.5 & 244\\
- & V  & 58053.1 & 58143.9 &  243 & 4.5 $\pm$ 0.6 & 3.8 & 5.4 & 243\\

& & & & & &  & & \\

Zowada &  B$_z$ & 58033.5 & 58208.2 &  74 & 1 &  0.77 & 1.07 & 34\\
- &  G$_z$  & 58028.5& 58197.2 &  73 & 1 &  0.81 & 1.06 & 36\\
-  & R$_z$ & 58028.5 &  58208.2 &  91 &1 & 0.9 & 1.04 & 37 \\
& & & & & &  & & \\

LCO+LT & g  & 58034.4 &  58257 & 309 & 5.8 $\pm$ 0.6&  4.6 & 6.4 & 156\\
- & V & 58034.4   & 58248.2 &  170  & 4.5 $\pm$ 0.6 & 3.7  & 5.6 & 80 \\
- & r & 58034.4 &  58257 &  435 & 5.4 $\pm$ 0.5 & 4.7  & 5.7 & 238 \\
- & i & 58034.4 &  58257 &  286 & 2.8 $\pm$ 0.3 & 2.4  & 3.1 & 154 \\
- & z & 58034.4 &  58257 &  280 & 2.2 $\pm$ 0.1 & 1.9  & 2.3 & 150 \\
& & & & & &  & & \\

\end{tabular}
    \label{tab:campaign}
\end{table*}{}

\subsection{Swift}
\label{Swift}
\textit{Swift} observed Mrk~110 three times per day from 26 October 2017 to 25 January 2018. Each observation totalled approximately 1~ks although observations were often split into two, or sometimes more, visits. The \textit{Swift} X-ray observations were  made by the X-ray Telescope \cite[XRT,][]{burrows2005} and UV and optical observations were made by the UV and Optical Telescope \cite[UVOT,][]{roming2005}. In total, 253 visits satisfying standard good time criteria,
such as rejecting data when the source was located on known bad pixels, 
(e.g. see \url{https://Swift.gsfc.nasa.gov/analysis/}
\url{xrt_swguide_v1_2.pdf}),  were made. 
The XRT observations were carried out in photon-counting (PC) mode and the UVOT observations were carried out in image mode. X-ray lightcurves in a variety of energy bands were produced using
the Leicester \textit{Swift} Analysis system \citep{evans2007}.
We made flux measurements for each visit, i.e. `snapshot' binning, thus providing the best available time resolution.  X-ray data are corrected for the effects of vignetting and aperture losses.

During each X-ray observation, measurements were made in all 6 UVOT filters using the 0x30ed mode which provides exposure ratios, for the UVW2, UVM2, UVW1, U, B, and V bands of 4:3:2:1:1:1.  UVOT lightcurves with the same time resolution were made 
using a system developed by \cite{gelbord2015}.
This system includes a detailed comparison of UVOT `drop out' regions, as first discussed in observations of NGC~5548 \citep{edelson2015}. When the target source is located in such regions the UVW2 count rate is typically 10-15\% lower than in other parts of the detector. The drop in count rate is wavelength dependent, being greatest in the UVW2 band and least in the V-band. The new drop-out box regions are based on intensive \textit{Swift} observations of three AGN, i.e. NGC~5548 \citep{edelson2015}, NGC~4151 \citep{edelson2017} and NGC~4593 \citep{mchardy2018}.  Observations falling in drop-out regions were rejected. 

\begin{figure*} 
\centering
\includegraphics[width=1.58\columnwidth]{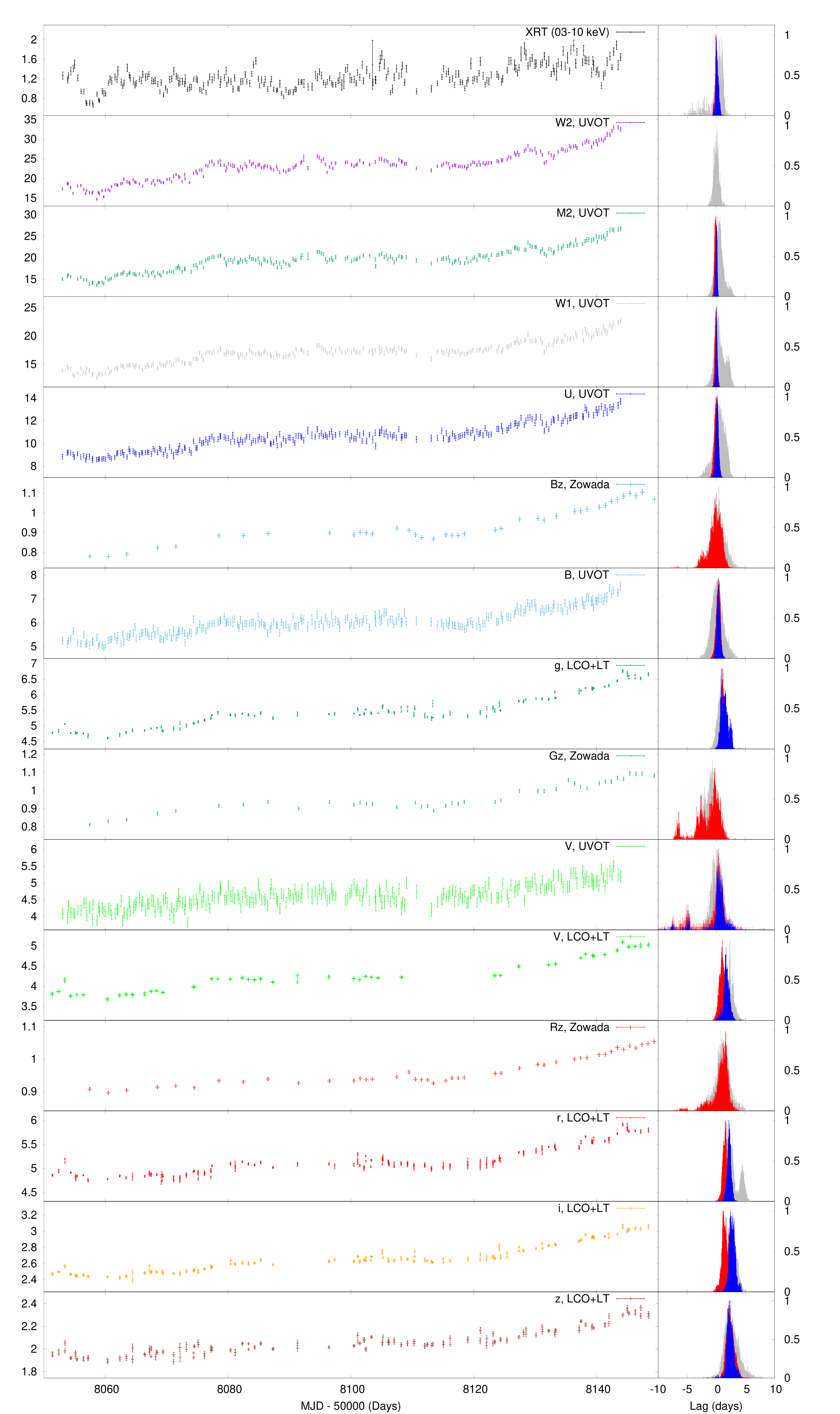}
\caption{ \textit{Left panels:} Lightcurves used for the short timescale lag analysis.   While the XRT lightcurve is shown in count s$^{-1}$, the measurements in the UVOT  filters  are in  ($\times 10^{-15}$)  erg cm$^{-2}$ s$^{-1}$ \si{\angstrom}$^{-1}$.  Regarding the ground-based observations,   LCO+LT  is in erg cm$^{-2}$ s$^{-1}$ \si{\angstrom}$^{-1}$, while RGB Zowada lightcurves are in arbitrary units. \textit{Right panels:} Lag vs UVW2 band probability distribution using unfiltered lightcurve (grey), and filtered ones by removing the linear trend (red) and a 10-day box-car moving average (blue).  } 
\label{fig:lc_short}
\end{figure*}

The resultant \textit{Swift} light curves, together with ground based observations described below (Section~\ref{groundbased}), are shown in Fig.~\ref{fig:lc_short}.
As observed previously \citep{bischoff1998,kollatschny2001}, Mrk~110 varied significantly in all bands during these observations (see $F_{var}$ in Table 2).

\subsection{Ground-Based Observations}
\label{groundbased}

Ground-based observations were performed with almost daily cadence between 2017 October and 2018 May 19 at LCO, the LT, and {\it Zowada Observatory}. Table \ref{tab:campaign} lists the start and end dates and number of visits for each filter.

\textit{Las Cumbres Observatory}: We used the 1 m robotic telescopes of the LCO network \citep{brown2013} and Sinistro imaging cameras to observe Mrk~110\footnote{Observations were taken under the LCO Key Projects KEY2014A-002 (PI Horne) and KEY2018B-001 (PI Edelson).}. Observations were taken in Johnson \emph{V}, SDSS \emph{g}$^{\prime}$\emph{r}$^{\prime}$\emph{i}$^{\prime}$, and Pan-STARRS $z_s$ filters (abbreviated hereinafter as \emph{griz} for convenience). Exposure times were typically $2\times300$ in the $V$ and $z$ bands, and $2\times180$ s in $g$, $r$, and $i$.

\textit{Liverpool Telescope}: Observations were carried out at the 2 m LT \citep{steele2004} using the IO:O imaging camera and \emph{griz} filters. Exposure times were $2\times10$ s in each filter.

\textit{Zowada Observatory:}  The {\it Zowada Observatory}, located in New Mexico, operates a 20-inch robotic telescope\footnote{\url{https://clas.wayne.edu/physics/research/zowada-observatory}}. During the Mrk~110 campaign it was in its full first season of operations and was equipped with \emph{RGB} astrophotography filters having effective wavelengths of 6365 \AA, 5318 \AA, and 4519 \AA\ respectively; these will be listed as $R_z$, $G_z$, and $B_z$ to distinguish them from other filter systems. Typically seven exposures of 100s were obtained per filter on each visit. 

\subsection{Ground-based Lightcurve Measurements}

All images were processed with standard methods as part of the observatory pipelines, including overscan subtraction, flat-fielding, and world coordinate system solution. Measurement of light curves from the LCO and LT data was carried out using the automated aperture photometry pipeline described by \citet{pei2014}. This procedure, written in IDL, is based on the photometry routines in the IDL Astronomy User's Library \citep{landsman1993}. The routine automatically identifies the AGN and a set of comparison stars in each image by their coordinates and carries out aperture photometry in magnitudes. An aperture radius of 4\arcsec ~was used, along with a background sky annulus spanning 10-20\arcsec. The comparison star data is then used to normalize the magnitude scale of each image to a common scale. For Mrk~110, 10 comparison stars were used. Measurements taken at a given telescope on the same night were averaged together to produce a single data point.

Photometric uncertainties include the statistical uncertainties from photon counts and background noise, and an additional systematic term determined by measurement of the excess variance of the comparison star light curves after normalization. The systematic term, measured separately for each telescope and each filter, accounts for additional error sources such as flat-fielding errors or point-spread function variations across the field of view. The uncertainties on each data point were combined as $\sigma^2_\mathrm{total} = \sigma^2_\mathrm{stat} + \sigma^2_\mathrm{sys}$. 

To account for small calibration differences between the telescopes, the LT light curves were shifted by adding a constant offset in magnitudes to bring them into best overall agreement with the LCO data. The magnitude scale for each filter was calibrated using comparison star magnitudes taken from the APASS catalog \citep{henden2012}. Finally, the light curves were converted from magnitudes to flux density ($f_\lambda$).

For the Zowada data, aperture photometry was performed using a circular aperture of radius 4 pixels (5.7\arcsec) on Mrk~110 and three nearby comparison stars.  The mean seeing FWHM was approximately 2.3 pixels.  The comparison stars were between 1--3 times brighter than Mrk~110.  Relative photometry was calculated from individual exposures, and the mean of all exposures for a given night used to create the AGN light curve in each filter. Since standard star calibrations are not readily available for the Zowada filter passbands, the flux scale of the Zowada lightcurves is not calibrated to an absolute scale. Without a flux calibration, the Zowada data can still be used for lag measurements but not for measurement of the AGN spectral energy distribution.
%As the data had observations collected from different telescopes, we combined the lightcurves in the same filter through the inter-calibration file software \textsc{CALI} \citep{li2014}. In particular the bands where this was applied were B (UVOT+Zowada), V (LT+LCO+UVOT+Zowada) and R (LT+LCO+Zowada). For this last band we combined r and R. Reference star used for the Zowada filters did not have any cataloged flux: the calibration was therefore performed with the time series renormalized  to their mean value. The obtained lightcurves were then simply rescaled to the average flux during the observation.

\section{Analysis}

\subsection{Variability: Intra-band variability and lags}

\subsubsection{X-ray vs UVW2 correlation}

\begin{figure*}
\includegraphics[width=\columnwidth,angle=0]{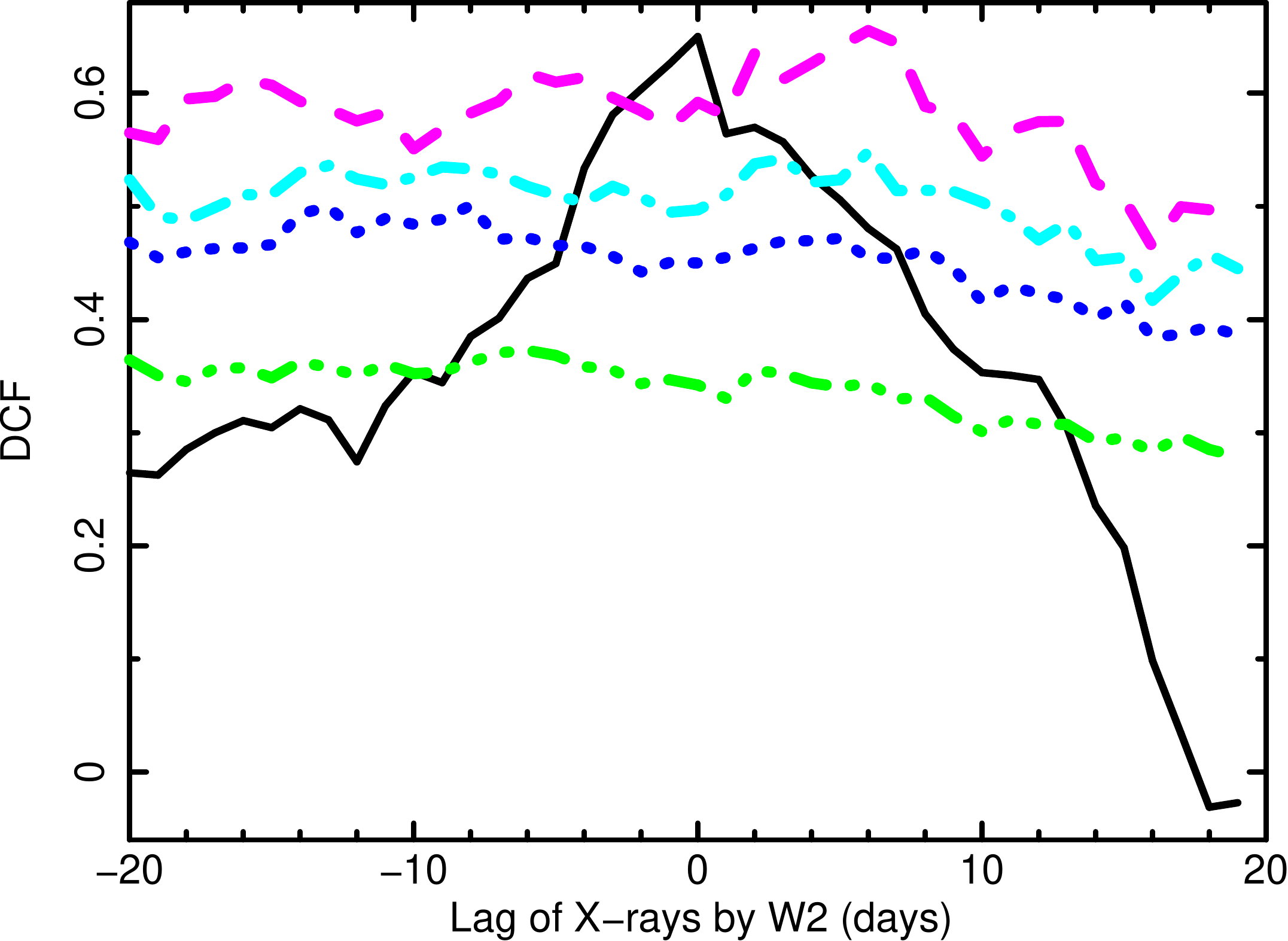}
\includegraphics[width=\columnwidth,angle=0]{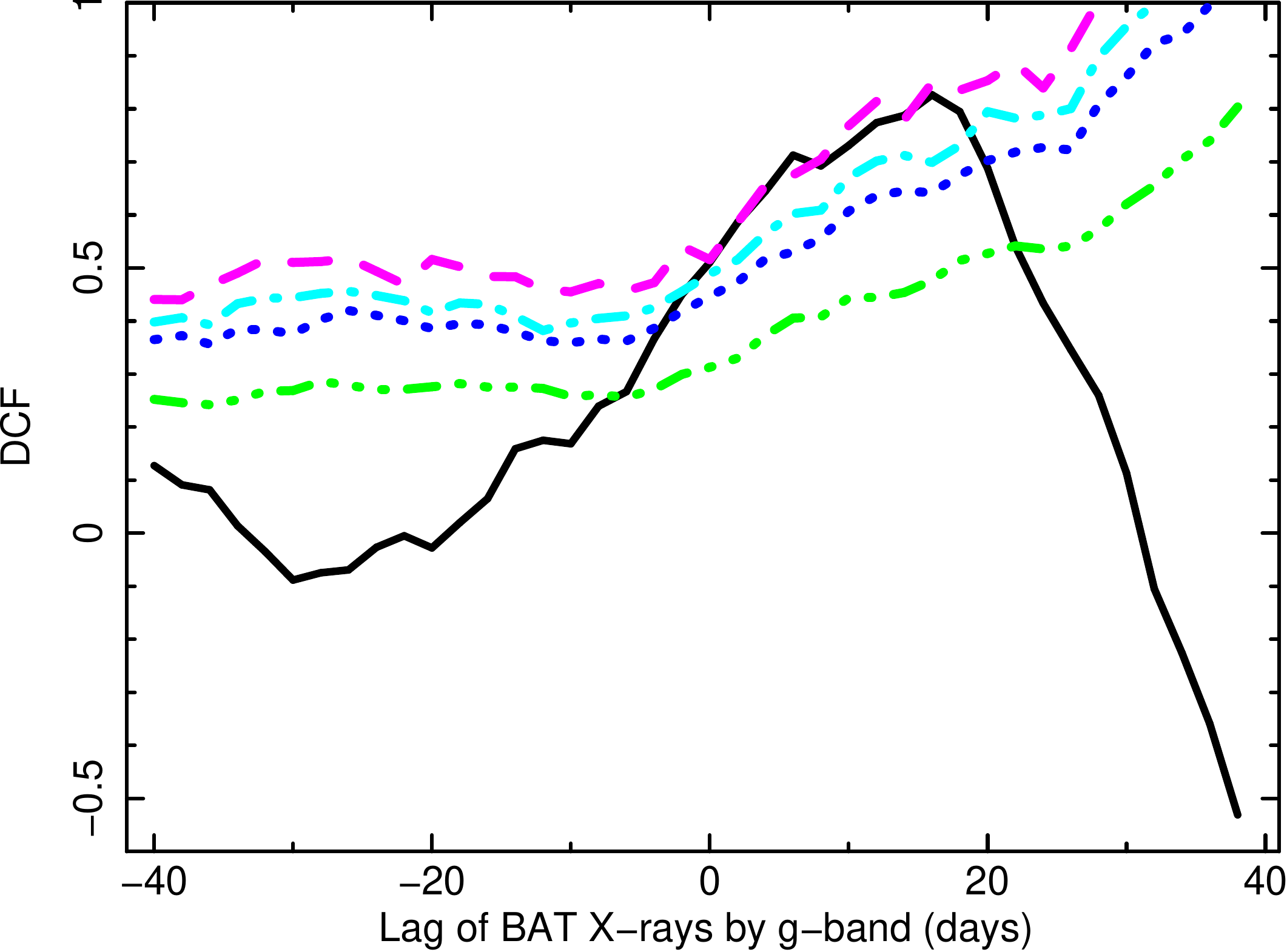}

\caption{\textit{Left Panel: }Discrete cross-correlation function (DCF) between XRT (0.3-10 keV) and UVOT (UVW2) lightcurve (see Fig. \ref{fig:lc_short}). Green, Blue, cyan and magenta line corresponding to 68,90,95 and 99\% confidence contours. \textit{Right Panel:} DCF computed with  the smoothed lightcurves (averaged with a 10 day box-car filter) from BAT (15-50 keV) and g band (see Fig. \ref{fig:lc})}.

\label{fig:dcf}
\end{figure*}

As a first step, we quantified the correlation between the X-ray and UVW2 bands.  We show in Fig. \ref{fig:dcf} the discrete correlation function \citep[DCF;][]{edelson1988} between these two bands, together with simulation-based confidence contours \citep{breedt2009}. We see that a correlation exists between these two bands at a confidence level greater than 99\%. No filtering has been applied to the lightcurves to remove long-term trends which often distort DCFs. Here the significance level of these contours is reduced, following the method outlined in \citet{mchardy2018}, to take account of the range over which lags are investigated. The present lightcurve simulation method follows \citet{emmanoulopoulos2013}, with code available from \citet{connoly2015}, which takes account of the count rate probability density function as well as the power spectrum of the driving lightcurve, unlike the method of \citet{timmer_and_konig_1995} which can only produce Gaussian distributed
light curves. The present X-ray lightcurve is not of sufficient quality to determine the shape of the power spectrum;  we therefore fixed the broken power-law spectral parameters at those derived by \citet{Summons2008} from combined RXTE and XMM-Newton observations (i.e. $\alpha_{low}=1, \alpha_{high}=2.8$ and $\nu_{bend} = 1.7 \times 10^{-6}$ Hz). The level of the confidence contours does not depend greatly on the exact choice of these parameters and, for any reasonable choice, the peak of the DCF always  exceeds the 95 \% confidence level and usually exceeds the 99 \% level.

\subsubsection{Short timescales}
Given the presence of a significant correlation between the X-ray and the UV band, we evaluated the lag as a function of wavelength with respect to the UVW2 band (1928 \si{\angstrom}). As mentioned above, the lightcurves show clear evidence of a long trend over the whole duration of the campaign, which could affect the measured lag \citep{white1994,welsh1999}. We therefore performed the calculation by  separating  short and long timescales. The canonical approach to study the variability at different timescales consists of using Fourier domain analysis techniques \citep{vaughan2003,uttley2014}. However, due to the structure of the data, such methods cannot be applied. We therefore filtered out the long-term trend applying two different approaches: in one case we subtracted a linear trend fitted between MJD 58050 and 58150 for the \textit{Swift} data and MJD 58050 and 58200 (i.e. when the peak of the lightcurve is reached) for the ground-based data; in the second case, following the procedure used  in by \citet{mchardy2014,mchardy2018} we subtracted a moving averaged trend, with a box-car width of 10 days \citep[see also][]{pahari2020}. The second method is equivalent to evaluating the lag between the two signals after applying a \lq\lq sinc" filter to their power spectrum, i.e. removing  variability longer than 10 day \citep{vanderklis1988}. In particular we chose a 10 day width due to the presence of few day gaps in the data. It is important to recall that applying the same filter to all lightcurves prevents the distortion of the lags. We notice that the measured lags remained stable for reasonably small variations of the box-car's width (i.e. a few days).

In order to compute the lags we used the so called \lq\lq flux randomization (FR) and random subset selection (RSS) method" \citep{peterson1998,peterson2004}. This method evaluates the lag distribution of the interpolated cross-correlation functions (ICCF) by re-sampling and randomizing the values of the fluxes at different epochs N times.    Given the structure of the data we limited the range of the possible lags between $-10$ and $+10$ days.

\begin{figure*}
\centering
\begin{center}
\includegraphics[width=0.8\textwidth]{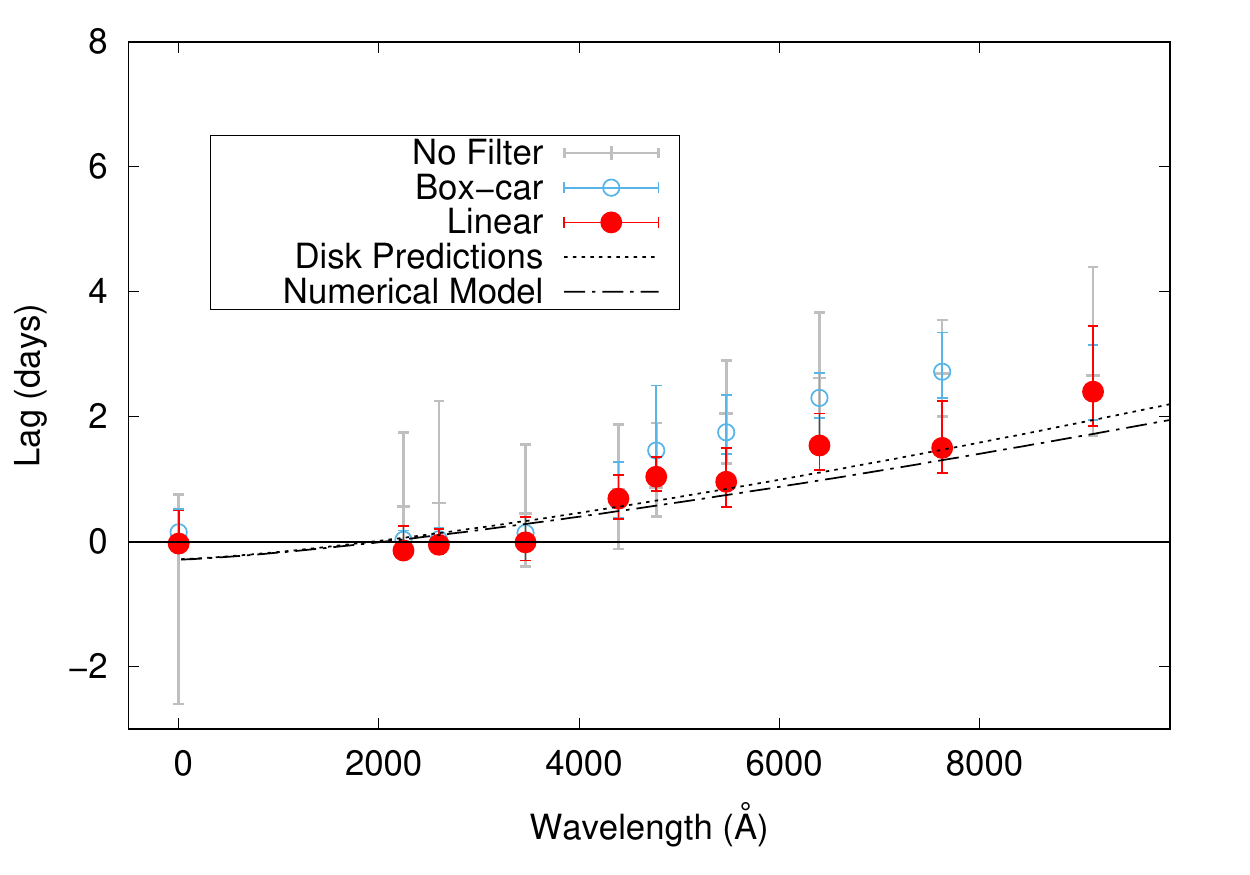}
\end{center}
\caption{Lag vs UVW2 band spectrum between MJD 58053 and 58143 using XRT, UVOT and LCO+LT data. For the grey points the lag was computed from the raw data, while for the full red and empty blue points data was filtered using linear interpolation and a box car filter respectively. Dotted and dot-dashed curve represent disc lag analytical and numerical predictions respectively. }
\label{fig:lagshort}
\end{figure*}

In Fig. \ref{fig:lc_short} we show the ICCF's  centroid probability distributions for the unfiltered and the two filtered  (box-car and linearly de-trended) lightcurves in all the bands. Given the small number of visits (and therefore the large gaps between the points), we did not apply a box-car filter to the Zowada data. The plots show that the distributions  obtained from the raw lightcurves are wider and often distorted; moreover it is also possible to see that the distributions computed with the linear de-trending method show almost always a shorter central lag.  The resulting lag-spectra are shown in Fig. \ref{fig:lagshort}, while the results for each band are reported in Table 2. The trends   show a clear evolution as a function of wavelength,  especially beyond 4000 \si{\angstrom}.  The values obtained between the two methods are consistent within the errors and at longer wavelength are smaller than the ones obtained in the non-filtered case,  suggesting the presence of a timescale-dependent time-lag.

In order to check the consistency of our result we also computed the wavelength dependent lag using the \textsc{JAVELIN} algorithm \citep{zu2011}, which has been recently shown to produce more realistic errors with respect to the FRRSS method \citep{edelson2019,yu2020}.   \textsc{JAVELIN} \textit{assumes} that the different lightcurves can be described as a damped random-walk \citep{zu2013}  linked by an impulse response function: therefore the lag is computed by constraining the parameters of the process and of the impulse response function through a Monte Carlo Markov-chain. The results for the raw lightcurve are shown in Tab. 2, and are fully consistent with the results obtained with the FRRSS.

\begin{table*}

\begin{adjustbox}{angle=90}
 \begin{tabular}{ l | c c |c c  | c c |c c  r }

Band & Wavelength (\si{\angstrom}) & Fvar (\%) & peak r & Lag  FRRS [days]   & peak r &  Lag FRRS  [days] &   peak r & Lag FRRS  [days] & Lag  Javelin [days] \\
 & & &  (Linear)   &  &  (Box-car) &  & (un-filtered) &   \\
\hline
     &&&&&&& \\

  X-ray & 0.25 & 17.6 & 0.29 $\pm$ 0.05 & -0.03$_{-0.12}^{+0.53}$ &  0.38$\pm$ 0.05  & 0.15$_{-0.17}^{+0.37}$ &  0.65 $\pm$      0.09  & -0.13$_{-2.47}^{+0.88}$ & -0.17$_{-0.04}^{+0.3}$\\ 
     &&&&&&& \\
    UVW2 & 1928  & 15.3&-  &   -& - &-  &-   & - &-\\
   &&&&&&& \\
  UVM2 & 2246 & 13.8 & 0.88 $\pm$ 0.02  & -0.14$_{-0.11}^{+0.39}$ &  0.77 $\pm$ 0.03 &  0.03$_{-0.13}^{+0.15}$ &  0.98 $\pm$      0.01 &0.56$_{-0.56}^{+1.19}$ & 0.01$_{-0.01}^{+0.03}$\\
     &&&&&&& \\
  UVW1 & 2600 & 11.5 & 0.86 $\pm$ 0.02  & -0.05$_{-0.15}^{+0.25}$ &   0.72 $\pm$   0.04&  0.04$_{-0.12}^{+0.18}$&  0.97 $\pm$      0.01& 0.62$_{-0.67}^{+1.63}$  & 0.01$_{-0.02}^{+0.05}$ \\
       &&&&&&& \\

  U & 3465 & 10.8 &  0.72 $\pm$ 0.03 & -0.01$_{-0.29}^{+0.41}$  &  0.57 $\pm$ 0.04 &  0.14$_{-0.24}^{+0.26}$& 0.95 $\pm$      0.01 & 0.45$_{-0.85}^{+1.10}$& 0.03$_{-0.16}^{+0.18}$\\
       &&&&&&& \\
  B & 4392 & 8.2 &  0.68 $\pm$  0.04 & 0.69$_{-0.33}^{+0.37}$  &  0.523 $\pm$ 0.05& 0.66$_{-0.28}^{+0.62}$ &  0.93 $\pm$      0.01 & 0.84$_{-0.96}^{+1.04}$ &  0.31$_{-0.38}^{+0.32}$\\
     &&&&&&& \\
     \hline
     
     &&&&&&& \\
     
  B$_z$ & 4500 & - &  0.68 $\pm$  0.04 &0.03$_{-0.78}^{+1.02}$ & - &- &  0.93 $\pm$      0.01 & $0.31_{-1.16}^{+1.34}$ & -\\
     &&&&&&& \\

  G$_z$ & r500 & - &  0.68 $\pm$  0.04 & -0.22$_{-1.33}^{+0.97}$  & - &- &  0.93 $\pm$      0.01 & -0.71$_{-0.59}^{+1.01}$ & -\\
     &&&&&&& \\

   R$_z$ & 6500 & - &  0.55  $\pm$ 0.05 & 0.99$_{-1.34}^{+0.76}$&  -  &  - &  0.94 $\pm$      0.01&  1.14$_{-1.09}^{+1.36}$  & - \\
 &&&&&&& \\

         \hline

     &&&&&&& \\

  g & 4770 & 14.1 &  0.7  $\pm$ 0.1 &  1.04$_{-0.24}^{+0.31}$   &   0.42 $\pm$ 0.11 & 1.46$_{-0.56}^{+1.04}$ & 0.93 $\pm$      0.03 &0.87$_{-0.47}^{+1.03}$ & 1.02$_{-0.09}^{+0.07}$\\
   &&&&&&& \\
  V & 5468 &    12.7 &  0.78  $\pm$ 0.03 & 0.96$_{-0.4}^{+0.5}$  &0.53 $\pm$    0.06 & 1.75$_{-0.35}^{+0.6}$ &  0.97 $\pm$      0.01& 2.05$_{-0.80}^{+0.85}$ & 2.16$_{-0.05}^{+0.84}$\\
     &&&&&&& \\
  r & 6400 & 9.4 &  0.55  $\pm$ 0.05 & $1.54_{-0.39}^{+0.51}$ &   0.5 $\pm$ 0.05  & 2.30$_{-0.32}^{+0.44}$ &  0.94 $\pm$      0.01& 2.79$_{-1.49}^{+1.27}$ &  3.04$_{-0.54}^{+0.03}$\\
 &&&&&&& \\

  i & 7625 & 10.7 &  0.54  $\pm$ 0.05 & 1.50$_{-0.40}^{+0.75}$   &  0.42 $\pm$ 0.05 & 2.72$_{-0.42}^{+0.63}$ &   0.91 $\pm$   0.01 &  2.69$_{-0.69}^{+0.86}$  & 2.29$_{-0.18}^{+0.52}$\\
 &&&&&&& \\
 z & 9132 & 9.4 &  0.51  $\pm$ 0.05 & 2.40$_{-0.50}^{+1.05}$  &  0.46 $\pm$ 0.06 & 2.39$_{-0.44}^{+0.76}$  & 0.91 $\pm$      0.01 & 2.66$_{-0.96}^{+1.74}$ & 2.09$_{-0.04}^{+0.06}$  \\
     &&&&&&& \\
     \hline
%&B$_{z}$  &   4500 &  9.7 &  0.79 $\pm$ 0.04 & -0.22$_{-1.33}^{+0.97}$& 0.47 $\pm$ 0.09 & 3.15$_{-1.50}^{+1.15}$& 0.97 $\pm$ 0.03& -1.11$_{-1.14}^{+0.91}$& 
%          &&&&&&& \\
%G$_{z}$ &   5500 &  9.7 & 0.74 $\pm$0.12 & -1.69$_{-1.61}^{+2.09}$ & 0.47 $\pm$ 0.09  &4.94$_{-2.09}^{+1.11}^\textbf{*}$& 0.95$\pm$ 0.01 & 0.31$_{-1.16}^{+1.34}$& 
%          &&&&&&& \\
%R$_{z}$ &   6500 &  6.9 & 0.65 $\pm$0.06 & 0.79$_{-2.54}^{+0.86}$& 0.47 $\pm$ 0.08  &5.50$_{-4.00}^{+0.80}^\textbf{*}$& 0.93$\pm$ 0.07 &1.05$_{-1.15}^{+1.40}$& 
    \label{tab:results}

\end{tabular}

    \end{adjustbox}
                \caption{Results of the lag vs UVW2 band computed between different bands. For sake of completeness we reported even the lags with multiple peaks, even though they are probably due to artifacts.}

\end{table*}{}

To better characterize the variability properties of the lightcurve  we also quantified the excess variance in each band and also their correlation coefficient with respect to the UVW2 band. As already seen in other AGN studies, the source showed a very strong correlation between the UVW2 and the longer wavelengths bands (all $\geq$ 0.9) and poorer correlation with the X-rays.  Repeating the same experiment using the filtered lightcurve, however, we notice two interesting features: first, the X-ray-UVW2 correlation becomes significantly poorer (0.29) and second, the correlation seems to decrease as a function of wavelength.  For the X-rays such a drop in correlation can be explained by the presence of much stronger fast variability in the X-rays (excess variance is almost 18\%).  On the other hand, at longer wavelength the poorer correlation is due to the decreasing variability  at longer wavelengths (see Tab. \ref{tab:campaign}).

%blue in ghz 6E+5 in gh 7E+14
%10^-19

\subsubsection{Long Timescales}

We analysed the overall variability during the campaign. In order to  better appreciate the long-term trend we smoothed the data with a moving average with a boxcar filter of 10 days width. Due to the different sampling strategies of the  various telescopes, we combined the lightcurves with similar filters. In particular, we merged  the V band data from UVOT and LCO+LT and the $r$ band data from LCO+LT together with the R$_z$ band from the Zowada observatory. To do this we used the inter-calibration software \textsc{cali} \citep{li2014}. Given that the  Zowada R$_z$ band is computed with  respect to the average, calibration was done in two steps: first, the $r$ band was also renormalized with  respect to its average before running the software, then the resulting lightcurve was multiplied by the average r band flux.

The ground-based lightcurves in the 4000 to 10000 \si{\angstrom} regime show a clear increase as a function of time reaching a peak around MJD 58200. However, XRT and UVOT covers only the first section of the lightcurve (see Tab \ref{tab:campaign}, Fig. \ref{fig:lc_overall} and \ref{fig:lc_short}). In order to cover also the peak of the lightcurve at higher energies we downloaded the BAT daily lighcurve from  \citet{krimm2013} from its online archive \footnote{ \url{https://swift.gsfc.nasa.gov/results/transients/}}. BAT lightcurves are known to show spurious variations due to the large field of view used with coded-mask technology. Even though the source detection is not significant at a 5 $\sigma$ level, the  variations observed in the data are present on timescales longer than $\approx$ a few days, and are not correlated with variations in the Crab or other nearby sources. Given also the strong resemblance between the BAT and the optical lightcurve, we conclude that the observed flare is most probably due to an intrinsic variation of the source and not a systematic effect.

\begin{figure*} 
\centering
\includegraphics[width=0.95\textwidth]{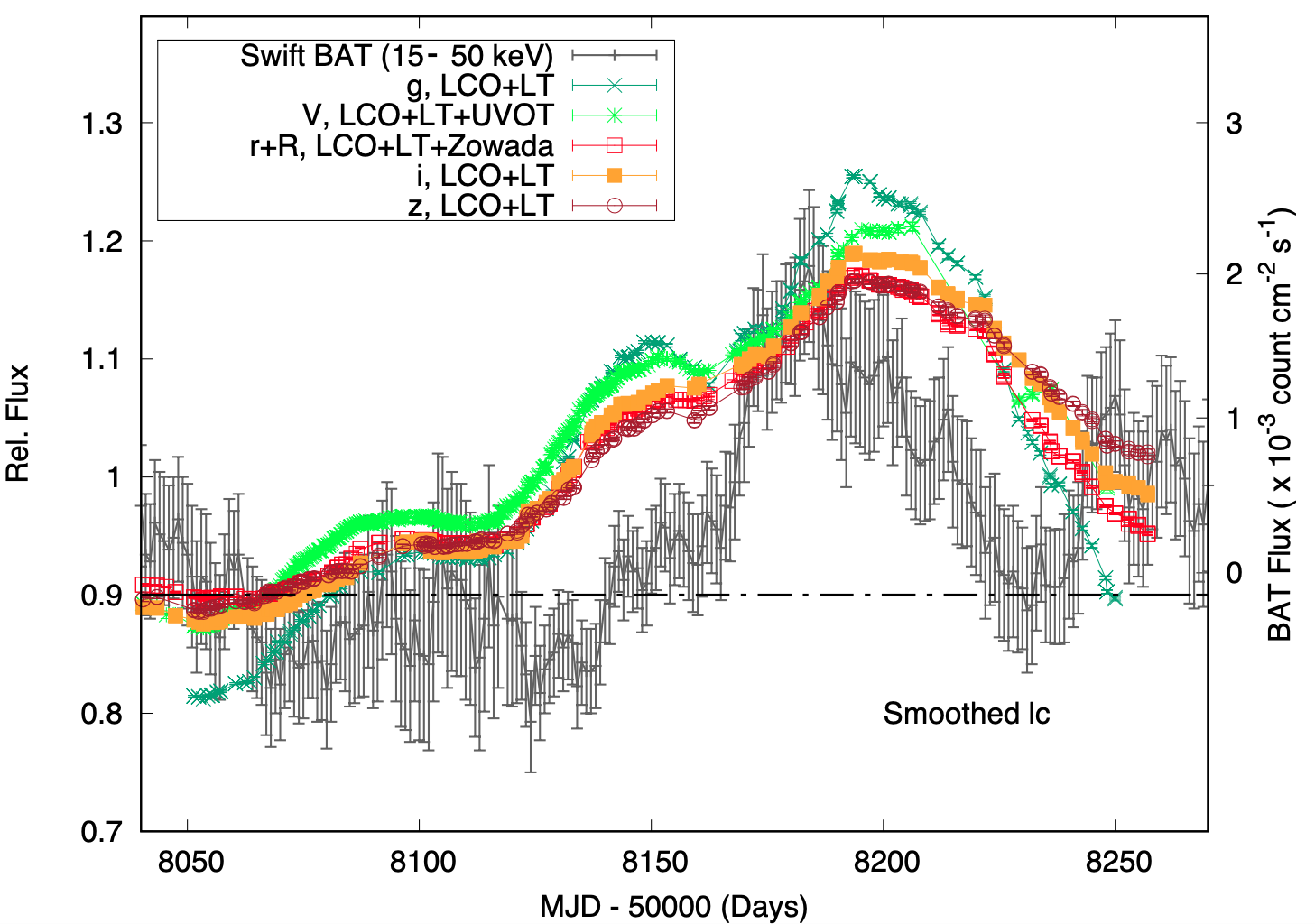}
\newline
\includegraphics[width=0.95\columnwidth]{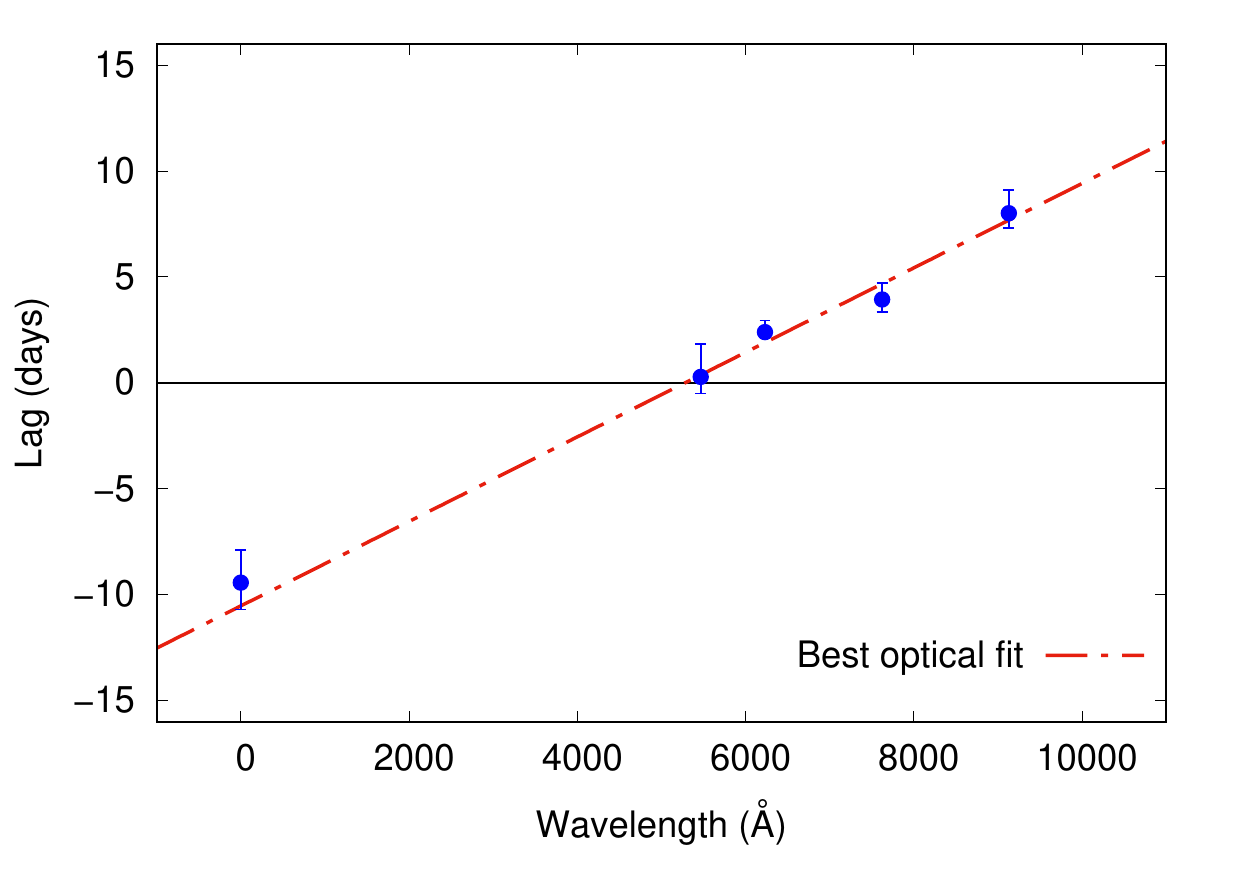}
\includegraphics[width=0.95\columnwidth]{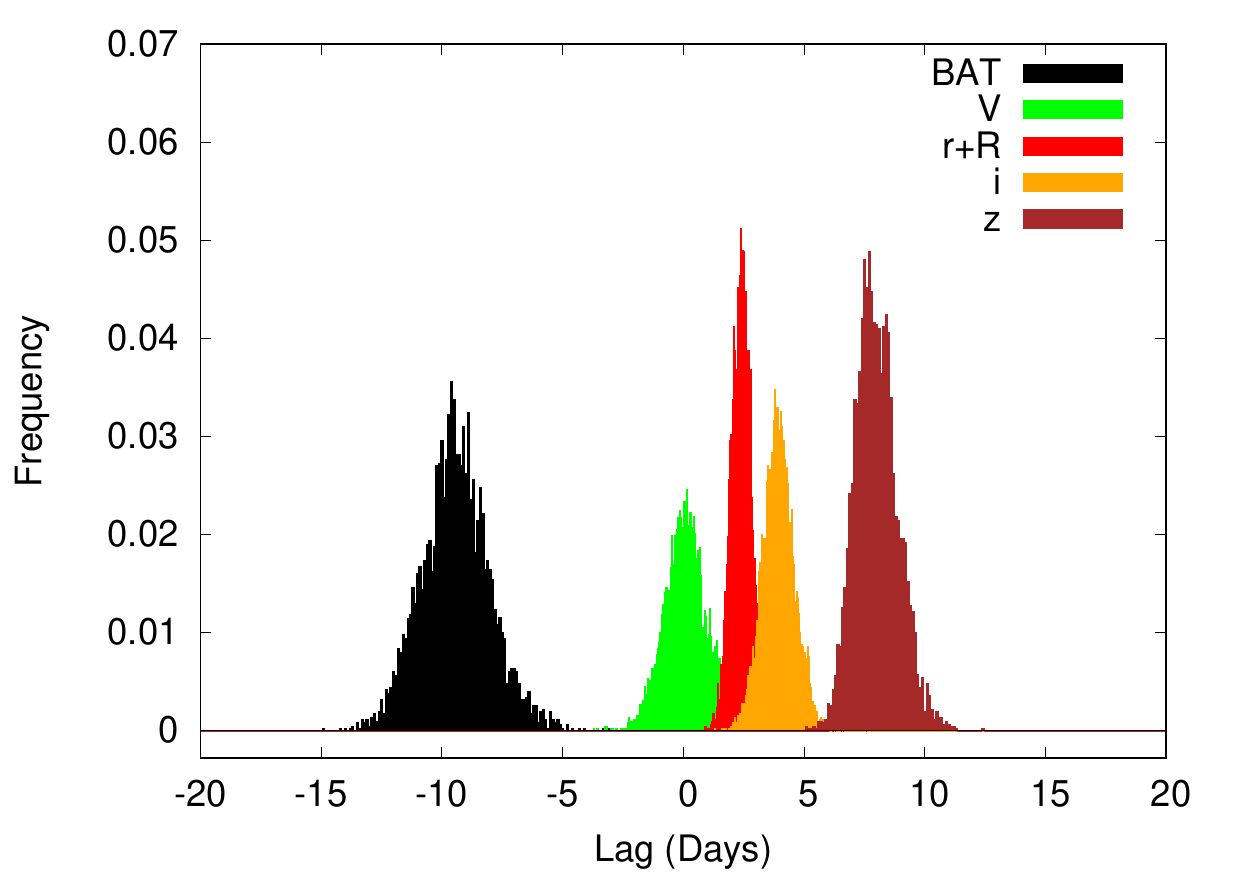}

\caption{  \textit{Top Panel}: Lightcurves used for the long-term trend analysis.  \textit{Bottom-Left Panel}: Lag vs wavelength  plot using long timescales.  \textit{Bottom-Right Panel}: Lag distributions from which the lag spectrum was obtained.}
\label{fig:lc}
\end{figure*}

Given the low statistics of the data, we  applied a 10-day box-car filter to smooth the lightcurve. In order to be able to compare properly its behaviour with the other filters, we applied the same procedure to the ground based ones. As shown in Fig. \ref{fig:lc}, a clear peak in the 15-50 keV band is seen at approximately MJD 58180. Moreover it is clear that while the rise of the BAT is significantly faster that the one observed at lower energy, the optical bands seem to decay with a slower rate as a function of wavelength.

 As in the previous section we computed the DCF to quantify the correlation between the BAT and the g  band, including the data between MJD 58140 and 58240. Sensibly small changes ($\approx$few days) in the choice of the dates and of the box-car width did not lead to any significant differences in our results.  Confidence contours  were derived in the same way as described in the previous section. Although the exact lag is not well defined by this simple DCF, the peak is around a lag of about 15 days and is significant at just below the 99 \% confidence level.

In particular, we use the same power spectral parameters to synthesise BAT lightcurves as are quoted for the earlier XRT lightcurve. Both BAT and long-term g-band have  here been smoothed, which will remove high frequency variability and should steepen the power spectrum.  The power spectrum for the smoothed BAT lightcurve is reasonably well fitted by a single power law of slope -1.5 but the bend frequency derived by the method of \citet{Summons2008} is within the range of the power spectrum so a model changing slope from -1 at low frequencies to -2.8 at higher frequencies could also be fitted. We have tried a variety of different underlying BAT power spectral shapes and the peak significance reaches the 99 \% confidence level.

\begin{table}
\begin{center}

\caption{Lag vs g band using the smoothed lightcurves: i.e. after applying a 10 day box-car moving average. Values are plotted in Fig. \ref{fig:lc}, bottom panels.}
\begin{tabular} { l | c  r  }
\hline

Band & Lag   [days]  \\
\hline
&\\
BAT &  -9.43 $_{-1.27}^{+1.53}$   \\
&\\
 V & 0.28$_{-0.78}^{+1.57}$  \\
&\\
r & 2.40$_{-0.78}^{+0.55}$   \\
&\\
i & 3.94$_{-0.59}^{+0.76}$    \\
&\\
z & 8.01$_{-0.70}^{+1.10}$  
\label{tab:long}
\end{tabular}
\end{center}

\end{table}

We then computed the lags as function of wavelength with the FRRS method  by using the 10-days smoothed lightcurves, and choosing the $g$ filter as reference band. The lag probability distributions are shown in  Fig. \ref{fig:lc} (bottom right panel). As expected  the $g$ is lagging behind the BAT curve by $\approx$10 days. The lag increases  as function of wavelength, going from  few days in the V band to almost 10 days  for the $g$ vs $z$ band (see Tab. \ref{tab:long} and Fig. \ref{fig:lc}, bottom panels). Given the shape of the lightcurves it is clear that the origin of the measured delay is the slower response of the longer wavelength during the decay after the peak. However it is interesting to notice that the lag spectrum for the long timescales seems to follow a linear trend from X-ray to near-IR (See Fig. \ref{fig:lc}).  We also tested the goodness of the lag by changing the width of the box-car, finding a stable lag between 5 and 13 days.  For shorter width the statistics becomes too low to measure a significant lag;  for larger box-cars, the long term trend is distorted by the smoothing, changing the value of the lag.

\subsection{Reverberation Modelling}

As a first unbiased approach we attempted to model and reproduce the variable emission from an X-ray corona illuminating the accretion disc through the MEMECHO algorithm \citep{horne1994} using all the collected lightcurves: i.e. attempting to reconstruct the impulse response function of the system through a maximum-entropy solution method \citep[see also][]{mchardy2018}.    We chose the X-rays as a reference band.  The algorithm manages  to reconstruct the lightcurve at lower energies, with an acceptable chi square ($\chi^2/N=1.1$). The obtained responses show clearly a broadening as a function of wavelength. However, it is also interesting to notice that almost all responses (especially at longer wavelengths) require a long tail which extends to $\approx$ 10-20 days, as seen also by the lag computed by using the long-term lightcurve. The excesses shown around 20 light days in the response of some filters are likely due to the presence of an X-ray dip around MJD 58060. The modelling using the UVW2 filter as a reference band showed a smoother trend.  Nevertheless the consistency of the shape of the response function among the different filters indicates that the presence of an extra component acting at longer timescales. 

\begin{figure}

\includegraphics[width=0.9\columnwidth]{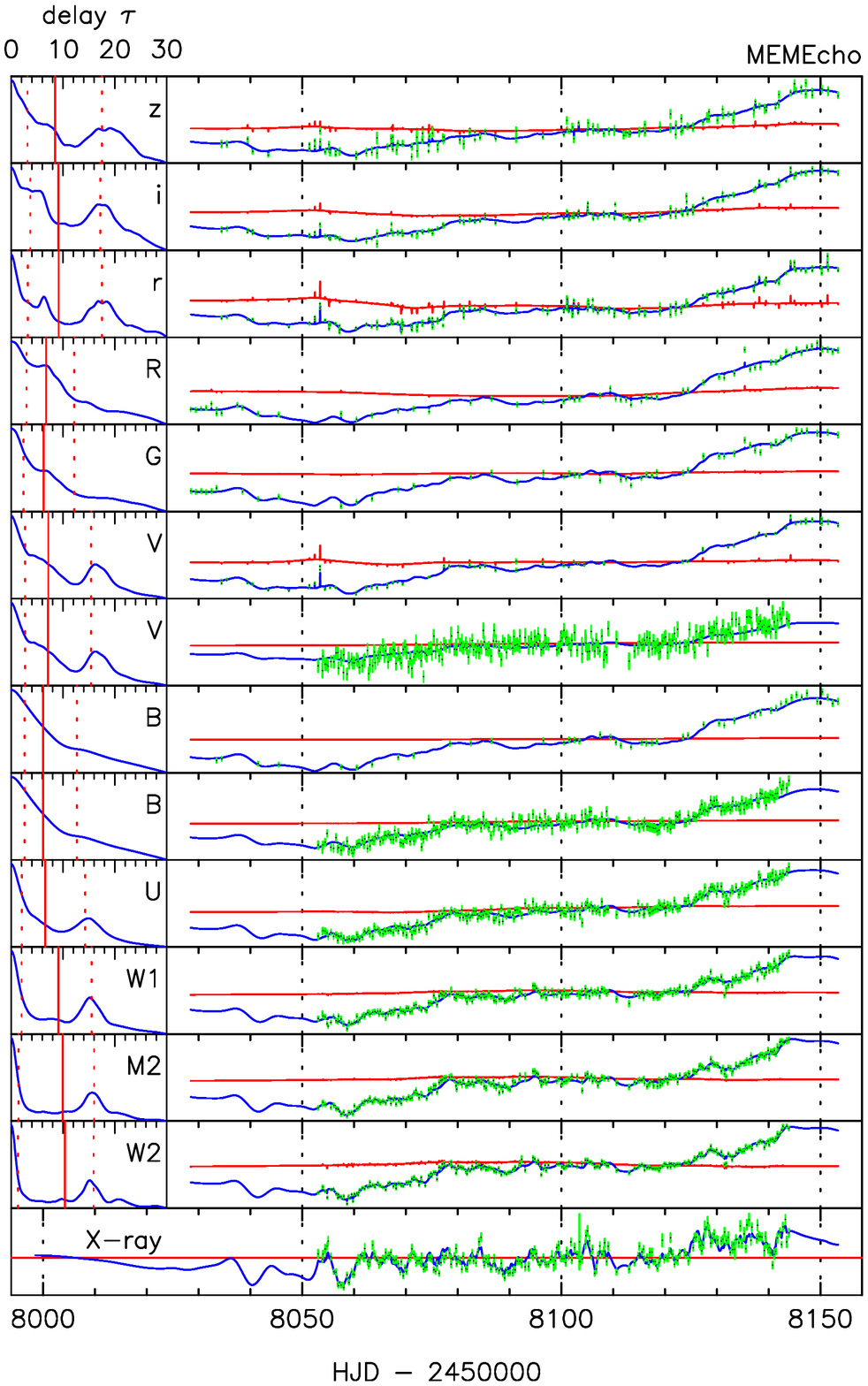}
\caption{Results from the MEMEcho modelling. \textit{Left Panels:}  Inferred response functions to be applied to the X-ray band in order to reconstruct the lightcurves at lower energies. \textit{Right panels:} Observed (green points) and modelled lightcurves (blue line). }
\label{fig:memecho}
\end{figure}

  The variable emission from an illuminated  accretion disc is expected to produce an increasing lag  ($\tau$) as a function of wavelength  ($\lambda$) as follows \citep{cackett2007}:

\begin{equation}
\centering
    \tau = \alpha \left[\left(\frac{\lambda}{\lambda_0}\right)^\beta-1 \right]
    \label{eq:fit}
\end{equation}{}

Where  $\lambda_0$ is the reference band wavelength (UVW2 band in this case: 1929 \si{\angstrom}) and $\beta$=4/3 for a \citet{shakura1973} accretion disc.   In our dataset we found two different components acting on different timescales. From the values of the lags and from the shape of the responses obtained by MEMEcho analysis, the short timescale component resembles the behaviour of an illuminated accretion disc. Given this we  perform a polynomial fit to the lag-spectrum obtained with the linearly de-trended lightcurve with Eq. \ref{eq:fit}  (fixing $\beta$=4/3) and obtained $\alpha = 0.28\pm 0.04 $ days with a reduced $\chi^2$ of 0.63. The low value is due to the large errors. We also fitted the lag leaving the power law index ($\beta$) as a free parameters and found  $\alpha = 0.1\pm 0.06$ days and $\beta=2.1\pm0.4$. The addition of  a free parameter decreases the reduced $\chi^2$ (0.44). This is mainly due to the  very short lag measured  below 4000 \si{\angstrom}. Even though the best slope is marginally steeper, the small $\chi^2$ and the large errors do not allow us to determine a significant deviation from the predictions of a standard accretion disc. %We therefore conclude that the  short timescales variability is most probably originating from an illuminated accretion disc.
%Following eq.  in \citep{fausnaugh2016}, and rescaling the mass,  we obtain an accretion-rate between 8 and 10 \%, so significantly smaller than past estimates.  A possible explanation for this however could be found in the high accretion-rate of this object, hence the temperature of the disc, which could dilute the observed lag. 
 
 We then performed a more detailed modelling using the same numerical code used by \citet{mchardy2018}. This model computes the response function of a  \citet{shakura1973} accretion disc illuminated by a  \lq\lq lamp-post" X-ray source (i.e. an point-source located over the rotation axis of the disc at a certain height). The expected lag is taken as half the time for the total light to be received.  We considered a mass of $2\times10^7$ M$_\odot$ \citep{bentz2015},  an accretion-rate  of $L/L_{Edd}$=40\% \citep{Meyer-Hofmeister}, a source height of 6 gravitational radii ($R_G$), and an inclination of 45$^\circ$\footnote{An inclination of 40$^\circ$ is usually taken as a standard  value for this class of sources \citep[][]{cackett2007,fausnaugh2016,kammoun,kammoun2021}. Recent X-ray spectral timing measurements of a similar source have shown evidence of a higher inclination \citep[][see however \citealt{caballerogarcia2020}]{alston2020}.  We notice  that while the inclination is known to have a strong effect on the X-ray spectrum, the dependence of the optical lags is expected to be negligble \citep{cackett2007,starkey2016,kammoun2021}.}. As shown in Fig. \ref{fig:lagshort}, even though the data points at shorter wavelengths seem to present a flatter trend, the predicted lag is in  good agreement with the observations.

\subsection{Spectral Energy Distribution and Energetics}

\subsubsection{Flux-Flux analysis}

In order to better characterize the origin of the variable emission, we performed a flux-flux analysis in order to separate the constant (galaxy) and variable (AGN) components following \citet{cackett2020} \citep[see also][]{cackett2007,mchardy2018}. We fitted the light curves using the following linear model:
\begin{equation}
f(\lambda, t) = A_\lambda(\lambda) + R_\lambda(\lambda)\cdot X(t) 
\label{eq:flux}
\end{equation}

\begin{figure*}
\centering
\begin{center}

\includegraphics[width=0.85\textwidth]{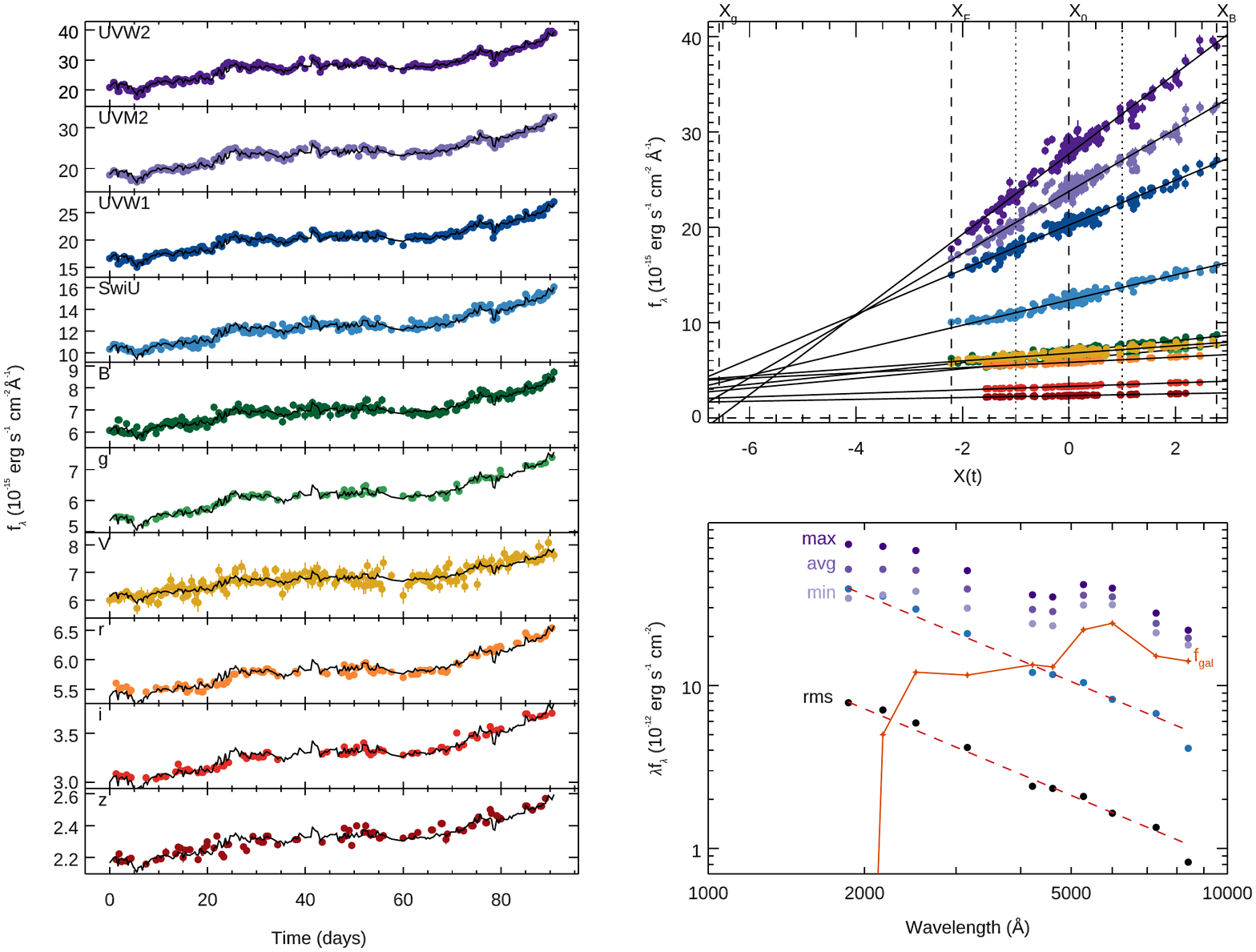}

\end{center}
\caption{Flux-flux analysis for the rising section  of the lightcurve. \textit{Left panels} shows the lightcurves, and the model (black line). The parametrisation described in Eq.~\ref{eq:flux}  represents the data well. \textit{Top right panel} shows the flux-flux relation between the different bands and the driving lightcurve X(t).  \textit{Bottom right panel} shows the optical/UV variable spectrum. Purple points on the higher section of the graph represent, minimum, average, and maximum values for each band. Orange continuous line represents the host galaxy  contribution. Blue points show the rms spectrum computed as maximum-minimum, while the black point represent the actual rms. Red dashed lines show the best fit.  }
\label{fig:fluxall}

\end{figure*}

Where X(t) is a dimensionless light curve with a mean of 0 and standard deviation of 1. $A_\lambda(\lambda)$ is a constant for each light curve, and $R_\lambda(\lambda)$ is the rms spectrum. All three parameters are determined by the fit. Such a simplified model does not take account of any time lags, adding some scatter around the linear flux-flux relations.  We first focused using the rise of the lightcurve during which all the facilities were observing. The results of the fit are shown in Fig. \ref{fig:fluxall}. We estimate the minimum host galaxy contribution in each band by extrapolating the best-fitting relations to where the first band crosses $f_\lambda$ = 0. The constant spectrum is consistent with the past host galaxy measurements by \citet{sakata2010}. The rms spectrum follows a  $\lambda F_\lambda\approx\lambda^{-4/3}$, as expected form the predictions of an accretion disc.

\begin{figure*}
\centering
\begin{center}

\includegraphics[width=0.8\textwidth]{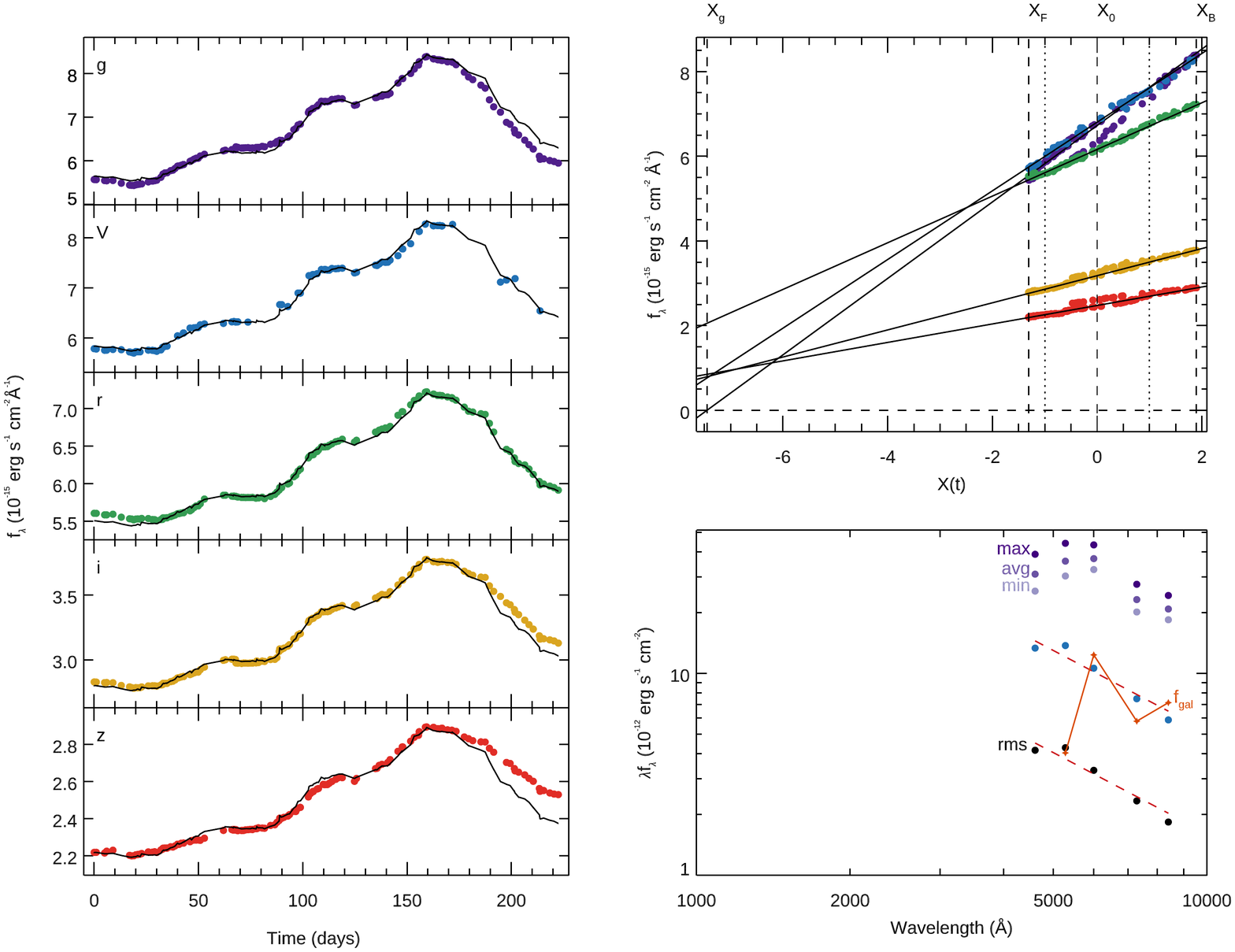}\vspace{-5mm}
\includegraphics[width=0.8\textwidth]{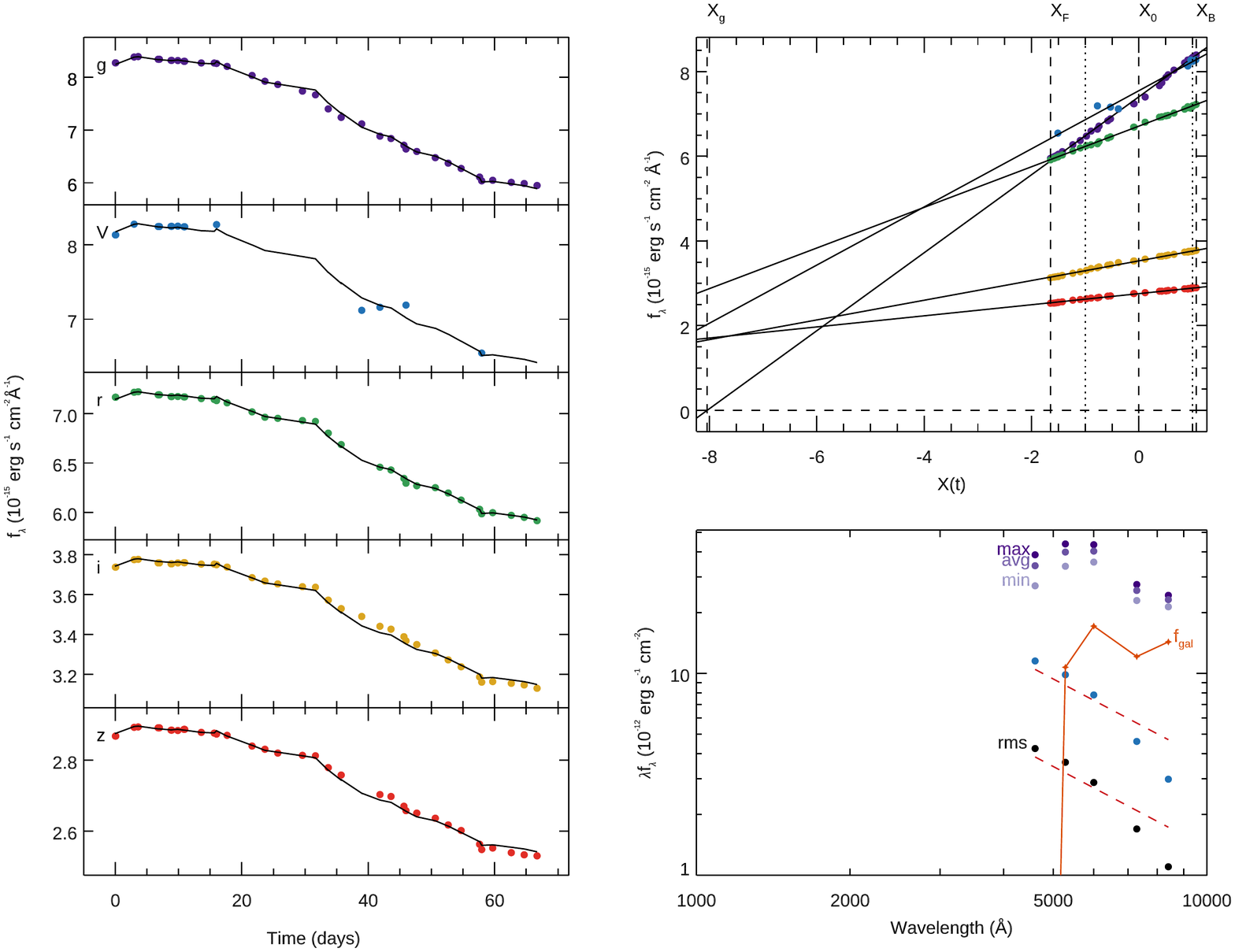}
\end{center}
\caption{ The two blocks show the flux-flux analysis for the whole lightcurve (Top) and only the tail (bottom). \textit{Left panels} shows the lightcurves, and the model (black line). An hysteresis is evident while using the whole lightcurve. \textit{Top right panels} shows the flux-flux relation between the different bands and the driving lightcurve X(t).  \textit{Bottom right panels} shows the optical/UV variable spectrum. Purple points on the higher section of the graph represents, maximum, average, and maximum values for each band. Orange continuous line represents the galactic contribution. Blue points show the rms spectrum computed as maximum-minimum, while the black point represent the actual rms. Red dashed lines show $\lambda f _\lambda \propto \lambda^{-4/3}$.  }
\label{fig:fluxsep}

\end{figure*}

We then analysed the long-term trend using the ground-based data. When plotting the modelled lightcurve with respect to the real observations, only the rise is well reproduced, while of the decay follows a different trend (see Fig. \ref{fig:fluxsep}, top panels).  We therefore analysed the tail of the long-term variation separately (see Fig. \ref{fig:fluxsep}, bottom panels), finding a significantly steeper spectrum ($\lambda F_\lambda\approx\lambda^{-2}$). The presence of a variable component different from a standard accretion disc can  explain the presence of a longer lag at longer timescales.

\subsubsection{SED: optical-UV/X-ray broadband}

\begin{figure*}
\includegraphics[width=0.9\columnwidth]{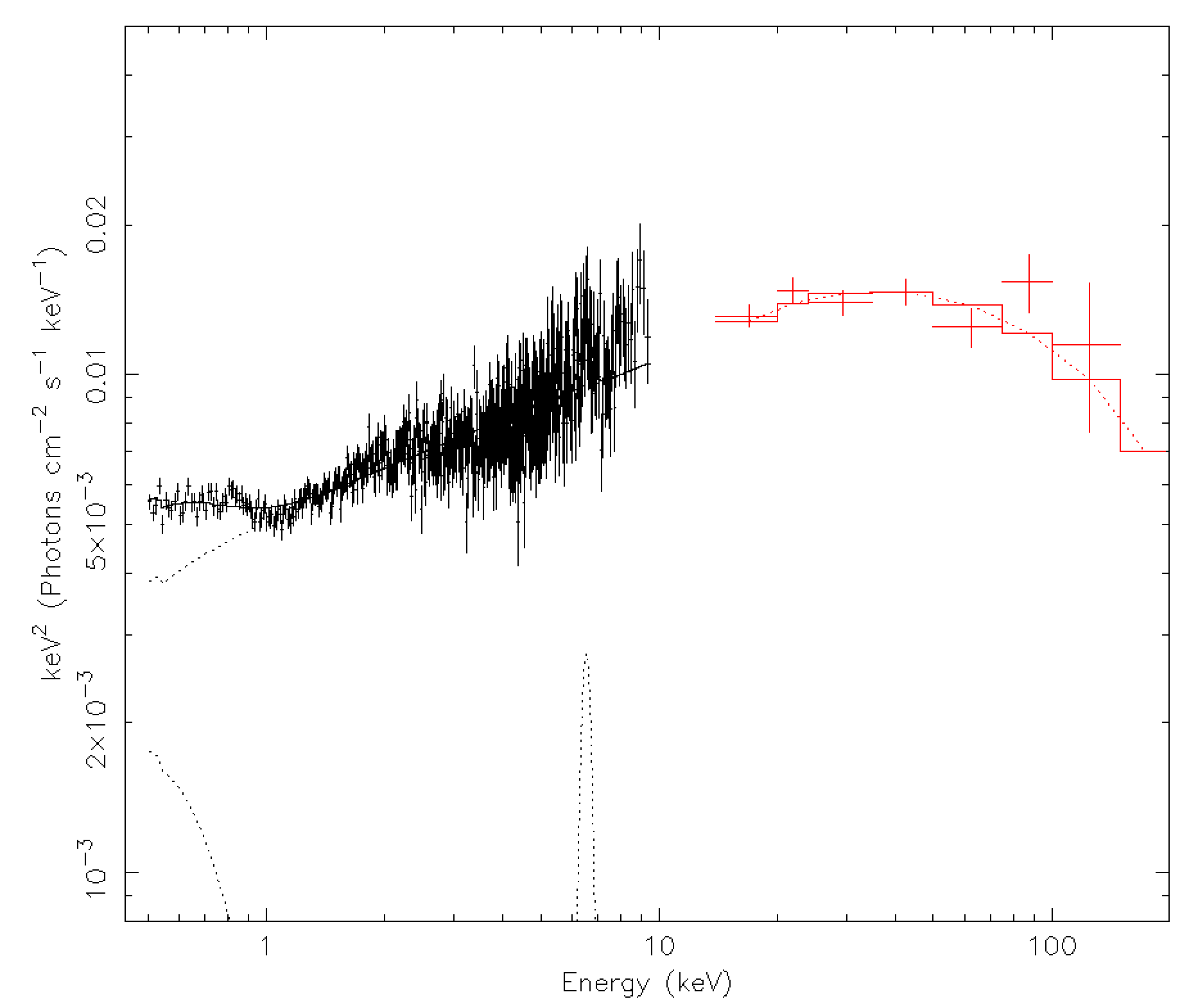}
\includegraphics[width=0.9\columnwidth]{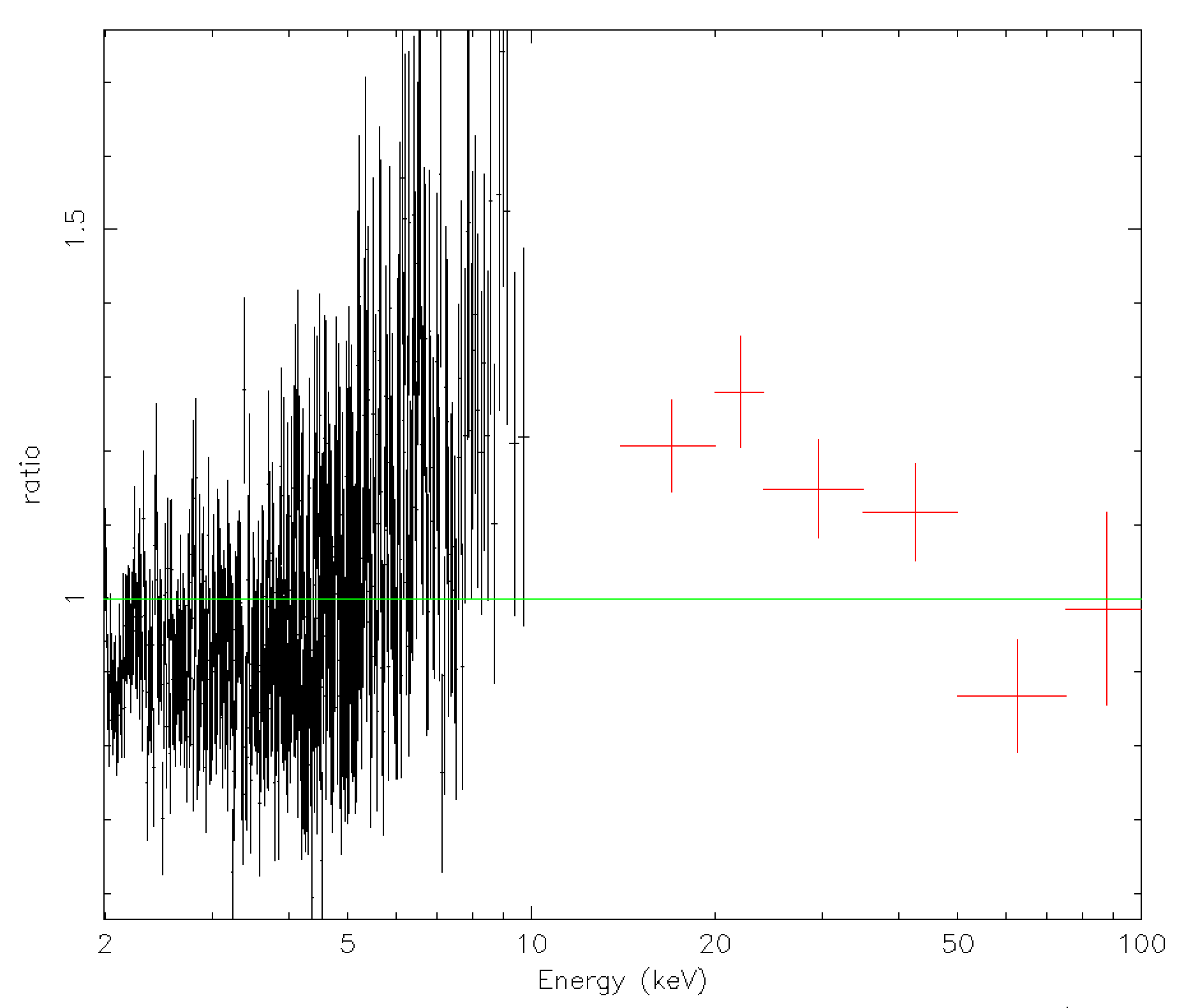}

\caption{\textit{Left Panel:} X-ray emission spectrum observed by \textit{\textit{Swift}} XRT (black points) and BAT (red points) and modelled with \textsc{Model 1}. \textit{Right Panel: } Ratio of the data to a simple power-law model. }
\label{fig:xspec}
\end{figure*}

The analysis reported so far shows  the presence of a clear connection between the X-ray and optical-UV variability, but does not take into account the energy budget of the two components. In order to quantify it, we estimated the variable flux in the different bands. We therefore built the X-ray spectral energy distribution from the XRT data including also the Mrk~110   BAT 70 month catalog spectrum. The flux in the 14-195 keV  range is  $5.7 \times 10 ^{-11}$ ergs cm$^{-2}$ s$^{-1}$ with  a power law slope ($\Gamma$) of 2.   The XRT data (0.3-10 keV), instead, shows a harder slope ($\Gamma_{3-5keV}=$1.66$\pm$ 0.07; see Fig. \ref{fig:xspec}, right panel). This has already been seen in past studies  \citep{idogaki} and suggests the presence of the so-called  \lq\lq Compton hump" \citep[see e.g.][]{guilbert_rees1988,lightman1988,george-fabian1991}. We therefore applied a simple model for an exponentially cut-off power law spectrum, reflected by neutral material \citep[i.e. \textsc{pexrav;}][]{Magdziarz1995}. We grouped the XRT spectrum with a minimum of 30 counts per bin using the FTOOL \textsc{grppha}.   We included in the fit a broad line for K$\alpha$ and the black-body emission for the soft excess. The results are reported in Tab. \ref{tab:spec} under \textsc{Model 1}. We obtained a fit with  $\chi^2$  of  439 (375 degrees of freedom). The model finds a power law slope  of $\Gamma\approx1.66 \pm0.02$ with a cut off at $\approx$ 120 keV and reflection fraction of 0.2. In order to estimate the total ionizing flux, we extrapolated the power-law flux in the 3-5 keV band to the 0.1-150 keV range: we obtained $F_{pl}=1\times 10 ^{-10}$ ergs s$^{-1}$ cm$^{-2}$.

In order to quantify the connection with the optical-UV emitting component it is necessary  to take into account the illuminating fraction.  Following the assumptions described in the previous section we estimate that total amount of radiation impacting the disc is $3  \times 10 ^{-11}$ ergs s$^{-1}$ cm$^{-2}$. Moreover, given that the source can vary up to 1 count s$^{-1}$ the variable flux we can take as an upper limit to the variable flux $\approx 2\times 10 ^{-11}$ ergs s$^{-1}$ cm$^{-2}$.

%We then  built the variable spectrum in the optical/UV range using the UVOT/LCO+LC data between MJD 8050 and 8140. We computed the variable flux as the difference between the maximum and the minimum flux in each band. The shape does not seem to follow a simple power law. This was already seen in past observations \citep{landt2011}, finding a strong excess for wavelength lower than 4000 \si{\angstrom}. We therefore performed a fit with a phenomenological broken power-law model. The spectrum follows a $\lambda^{4/3}$ for after 4000 \si{\angstrom}, in agreement with the results by \citep{landt2011}.    

From the rms-spectrum computed  in  the previous subsection, we obtained a variable optical-UV flux of $2.5\times 10 ^{-11}$ ergs s$^{-1}$ cm$^{-2}$.  We conclude therefore that  the X-ray variations contain enough energy to power most of the variability seen in the optical-UV (if they are driven by an illuminated accretion disc). We also notice that if the low energy  emission is dominated by the BLR, the solid angle to which it is exposed will be much larger, and  the argument will still be valid. 

The reported estimate for the ionizing variable flux sets a solid lower limit, which, however, is based on a purely phenomenological model. In order to have a more realistic value, we also fitted the SED using the physically motivated model \textsc{optxagnf} \citep{done2012}. The model \emph{assumes} the presence of emission from an extended hot corona, going from the soft X-rays to the far UV, while optical and near UV emission would be dominated by the accretion disc.

\begin{table*}
\caption{ Best fit parameters modelling. \textsc{Model} 1 was applied to the only X-ray data from XRT+BAT and is consists in \textsc{zphabs} $\times$ [\textsc{pexrav+zgauss+zbbody}].   Errors are reported with 90$\%$ confidence interval contour. In order to obtain the fit we froze the following parameters: nH=1.27 $\times$10$^{20}$ [cm$^{-2}$] , z = 0.035; He abund (elements heavier than He) = 1; Fe abund=1. {\textsc{Model} 2 includes also optical and UV wavelength, and was parametrized with using \textsc{zphabs} $\times$\textsc{optxagnf}. Mass was fixed to 2$\times10^7$M$_\odot$, accretion-rate  to 40$\%$, R$_{out}$=10$^3$R$_G$,  spin $a$  to  0}. }
\begin{center}
    
\begin{tabular} { l c r | l c  r  }
\hline

& Model 1  & &  & Model 2  & \\
&   \textsc{zphabs} $\times$ [\textsc{pexrav+zgauss+zbbody}]  & &  &  \textsc{zphabs} $\times$\textsc{optxagnf}  &\\

Parameter && Best Fit  & Parameter && Best Fit\\
\hline
 && & && \\
$\Gamma$ && 1.66$\pm$0.02  & $\Gamma$ && 1.72 $\pm$0.02\\
 && & && \\
E$_\mathrm{cut}$ [keV] && 118$_{-22}^{+33}$ & a && 0 \\
 && & &&   \\
refl$_\mathrm{frac}$ && 0.21$_{-15}^{+0.17}$ &  R$_\mathrm{cor}$ [R$_\mathrm{G}$] && 70$\pm$20\\
   && & && \\
norm$_\mathrm{pexrav}$ ($\times$10$^{-3}$) && 5.2$\pm$ 0.1 & kTe [keV] && 0.25$\pm$0.08\\
  && & &&  \\
Line$_\mathrm{E}$ [keV] && 6.7 $\pm$ 0.1 & $\tau$ && 11$\pm$2\\
  && & &&  \\
$\sigma_\mathrm{E}$ [keV] && 0.3$_{-0.2}^{+0.8}$ & f$_\mathrm{pow}$ && 0.5 $\pm$0.1\\
  && & &&  \\
norm$_\mathrm{zgauss}$ ($\times$10$^{-5}$) && 4.2$\pm$ 0.1 \\
  && & &&  \\
kT [keV] && 0.11$\pm0.05$\\
 && & && \\
norm$_\mathrm{zbbody}$ ($\times$10$^{-5}$) && 5.62$\pm$ 0.4\\
 && & && \\
 $\chi^2$/ d.of. & & 648 / 590 &  $\chi^2$ / d.of.  && 900/596 \\
\label{tab:spec}
\end{tabular}
\end{center}

\end{table*}

{We included optical and UV data average fluxes,  subtracting the galactic contribution. The fit was performed by fixing  mass accretion-rate $L/L_{Edd}$=40$\%$ and spin ($a$) to 0. Mass, redshift and distance were fixed accordingly to values reported in the introduction. The best fit model reproduced the data with an acceptable chi square ($\chi^2$ / d.o.f = 769 / 594; see Tab. 4). Some excess is found at higher energies and in the UVOT band. This means that the flux will be slightly overestimated.  We also attempted to fit the data  with $a=0.998$, but, despite values were still in the same range of uncertainties of the previous fit, the obtained $\chi^2$ was significantly worse ($\chi^2$/d.o.f.$\approx$2). However an acceptable $\chi^2$ was found for a mass of 2.5 $\times 10^7$M$_\odot$ \citep[which is still within the range of uncertainties for most f-factors][]{bentz2015}.  We estimated a total ionizing flux (5 eV -1000 keV) of  F$_{ion}\approx$ 6$\times$ 10$^{-10}$ erg cm$^{-2}$ s$^{-1}$ (i.e. L$_{ion}\approx$7$\times$ 10$^{45}$erg  s$^{-1}$).}

\subsubsection{HST spectroscopy}

The diffuse continuum emission from the BLR will be a source of wavelength-dependent lags in addition to lags from reprocessing in the accretion disc \citep[][]{korista2001,korista2019,lawther2018}.  In order to estimate the contribution from the BLR diffuse continuum  we obtained Hubble Space Telescope ({\it HST}) spectra of Mkn~110.  Observations were performed on 2017-12-25, 2018-01-03, and 2018-01-10 using STIS.  Observations were taken with the 52'' $\times$ 0.2'' aperture using the G230L, G430L and G750L gratings.  During each epoch exposures totally 800s (G230L), 180s (G430L and 180s (G750L) were obtained. In addition to the standard pipeline-processing we used the {\sc stis\_cti} package to apply Charge Transfer Inefficiency corrections to the data. Moreover, we used the contemporaneously obtained fringe flats to defringe the G750L spectra.  Any remaining large outliers in the spectrum that were identified visually as hot pixels were removed manually.

We fit the mean spectrum with a variety of models to estimate a plausible contribution from the BLR diffuse continuum.  To model the BLR diffuse continuum we use a model calculated by \citet{korista2019} that is tailored for NGC~5548.  This assumes a locally-optimally emitting cloud model for the BLR, with a fixed cloud hydrogen column density of $\log N_{\rm H} (cm^{-2}) = 23$, and a gas density distribution ranging from $\log n_{\rm H} (cm^{-3}) = 8 - 12$ (solid black line in Fig. 4 of \citet{korista2019}; see that paper for full details of the model calculation).  We Doppler-broaden the diffuse continuum model by convolving with a Gaussian with FWHM = 5000 km/s \citep[][]{peterson1998,peterson2004}, consistent with emission from the inner BLR in Mrk~110.  The presented diffuse continuum model does not include the unresolved high-order Balmer and Paschen emission lines redward of those jumps. In addition to the pile-up of Doppler-broadened higher-order Balmer, Paschen, and other emission lines (HeI and FeII), there is another effect that may serve to smooth the free-bound continuum emission jumps in wavelength space. The \textsc{cloudy} photoionization model simulations all assume that the wavelengths of the Balmer and Paschen series limits are at their low-density values, with these values then corrected for atmospheric refraction. However, due to the likely presence of high density gases within the BLRs of AGN (e.g., 10$^{12-13}$ cm$^{-3}$), the wavelengths of the free-bound jumps are likely shifted to somewhat longer wavelengths due to the finite sizes of the emitting hydrogen atoms. The contributions of mixtures of continuum emission from gas spanning the full range of expected densities within the BLR will thus smooth out somewhat the abrupt free-bound continuum jumps in wavelength space.  This effect is also currently not included in the model spectral template. As neither of these effects are included in the spectral template for the BLR diffuse continuum, we exclude the wavelength ranges 3600 -- 4050 \AA~and 8000 -- 8700 \AA~from the fit. 

We further note that we have not calculated a model specific to the line luminosities of Mkn~110, and, while the general shape of the diffuse continuum spectrum is common to all models, the amplitudes of the Balmer and Paschen jumps do depend somewhat on model assumptions.  For these reasons we view the present spectral fitting process as simply providing a guide and an estimate of the BLR diffuse continuum contributions to the spectrum.

For the underlying AGN continuum emission we try two different models - the first is a simple power-law, the second is a standard accretion disc model \citep[][]{shakura1973} with the outer radius set to be equivalent to a temperature of 2000K.  This produces a spectrum following the relation $F_\lambda \propto \lambda^{-7/3}$ through most of the wavelength range, but begins to rollover at the longest wavelengths.  These two models test the sensitivity of the results to assumptions about the accretion disc emission.  We include both UV and optical \ion{Fe}{II} templates.  In the optical we use the template of \citet[][]{veron-cetty}, while in the UV we use the model of  \citet[][]{mejia2017}. This latter model has the advantage that it covers a broader wavelength range in the UV than other models.  We investigate Doppler broadening in these \ion{Fe}{II} emission models by convolving with a Gaussian, and find best fits for FWHM = 3000 km/s. We fit broad and narrow emissions lines with Gaussians. For the host dust emission from the torus we include a single temperature blackbody with T=1800K, while this is simplistic \citep[e.g.  see Appendix A of][for a more complex model]{korista2019}, longer wavelength coverage is needed to better constrain this component.  We also include an 11 Gyr old solar metallicity stellar population model \citep[][]{bruzual2003}, broadened to the resolution of STIS G430L and G750L. We use the surface brightness profile fit parameters of \citep[][]{bentz2009} to scale the galaxy flux to the slit width (0.2\arcsec) and extraction size (0.36\arcsec) used for these {\it HST} data, finding 6.4$\times10^{-17}$ erg s$^{-1}$ cm$^{-2}$ \AA$^{-1}$ at 5100\AA.  This is fixed in all the fits.

\begin{figure*}
\centering
\begin{center}

\includegraphics[width=0.45\textwidth]{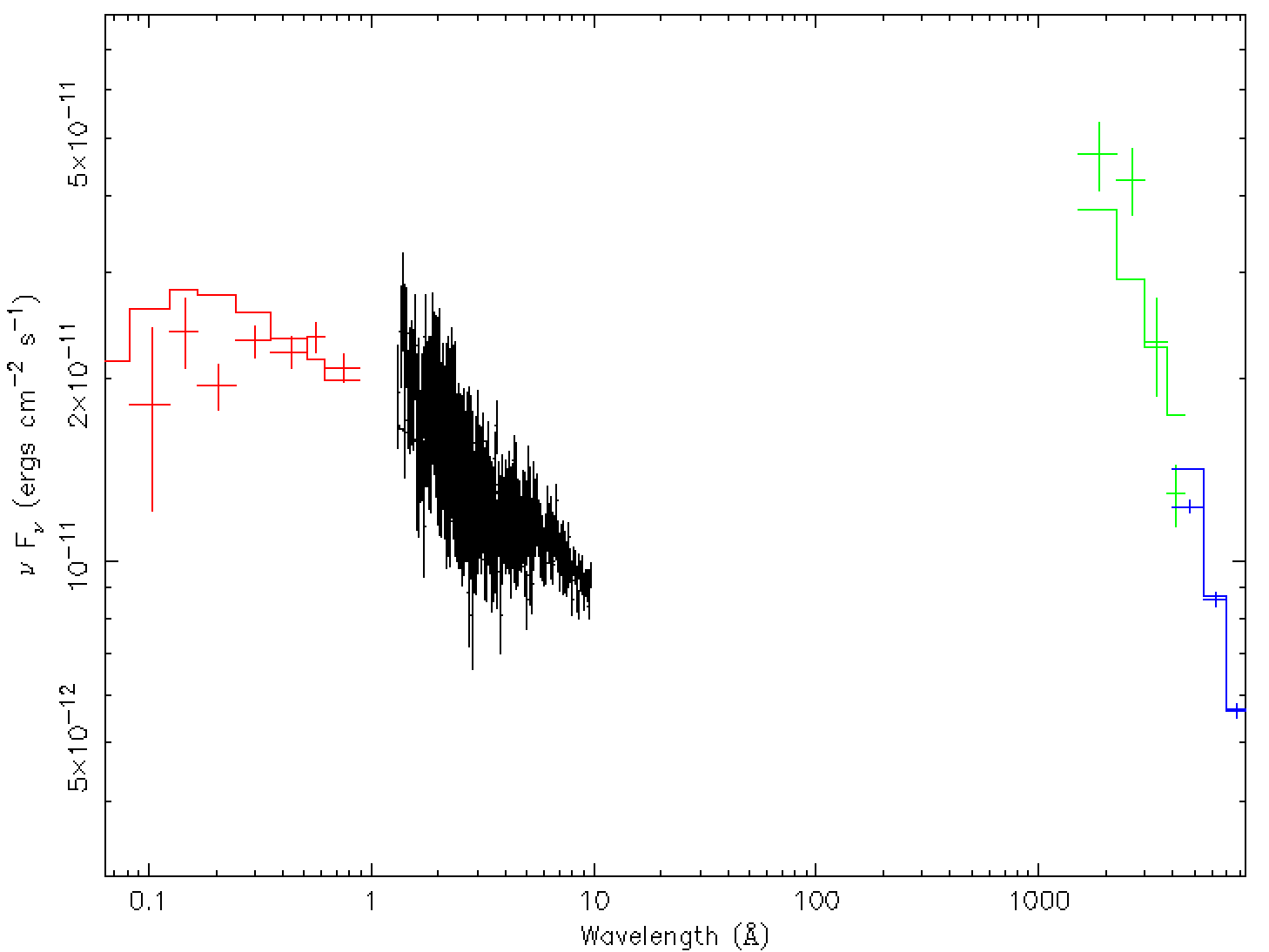}
\includegraphics[width=0.45\textwidth]{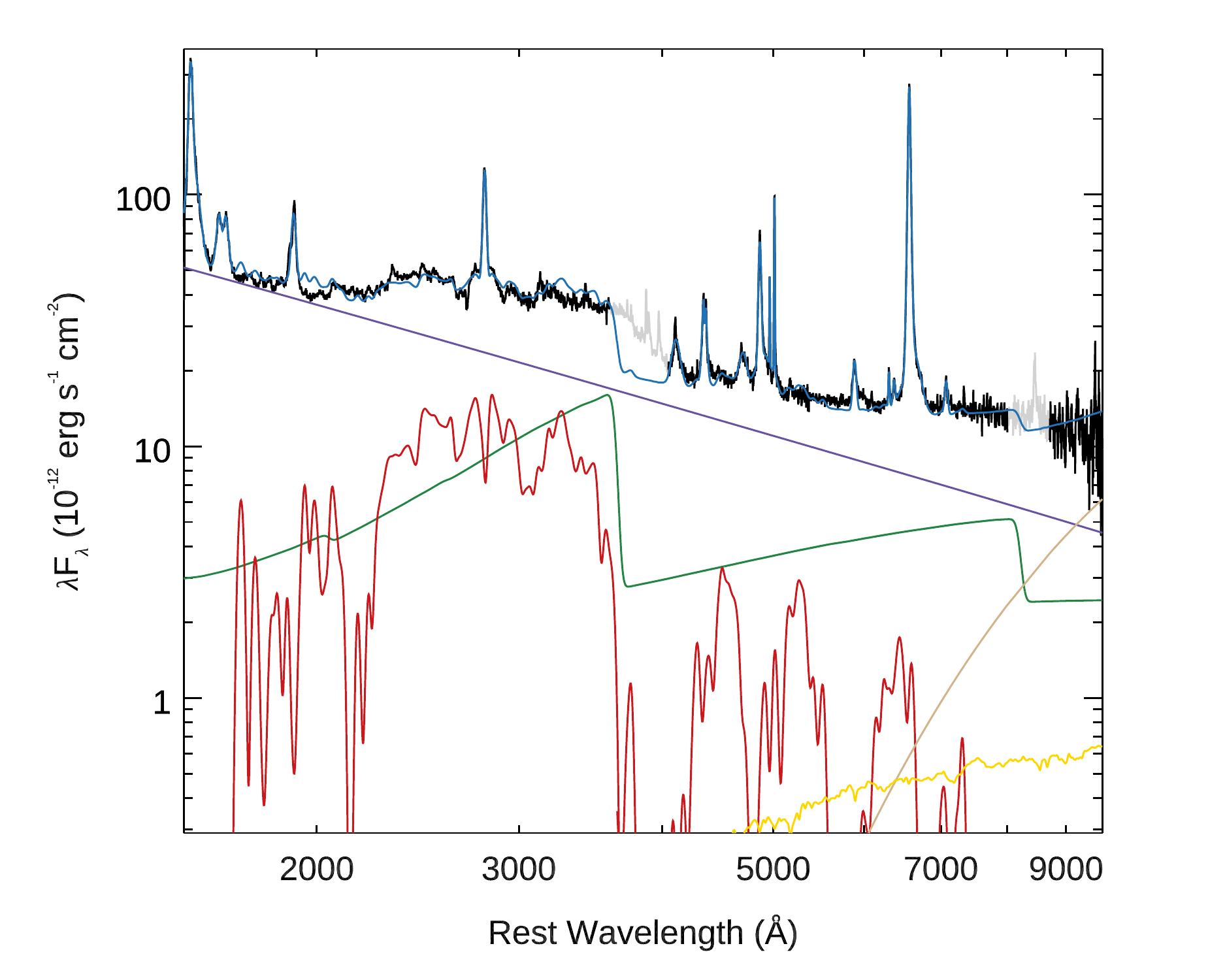}
\end{center}
\caption{\textit{Left Panel:} X-ray + optical/UV spectrum of Mrk~110 measure with \textit{\textit{Swift}} BAT (red points), XRT (black points), UVOT (green), LCO+LT (blue) data. Data was fitted with \textsc{Model 2}. \textit{Right Panel:} Mean HST spectrum of Mkn 110 (black).  The light gray line shows the wavelength regions excluded from the fit.  Blue is the overall model, while individual components shows are the accretion disc (purple), diffuse continuum (green), \ion{Fe}{II} template (red), stellar population template (yellow) and hot dust blackbody (tan). }
\label{fig:uvspec}

\end{figure*}

The diffuse continuum model described above is fitted including a scale factor allowing it to best-match the data.  Although we have not optimized the modelled diffuse continuum spectrum for the broad emission line spectrum of Mrk~110, we checked that the spectral fit's scaling of the diffuse continuum is in line with the model predictions for NGC~5548 by \citet[][]{korista2001,korista2019}.  The best-fitting model using an accretion disc is shown in Fig. \ref{fig:uvspec}, and generally does a reasonable job of fitting the overall shape of the spectrum from $\sim$1600 -- 10000\AA.

We perform synthetic photometry on the best-fitting power-law and disc models to estimate the fraction of the flux contributed by each component in each of the {\it Swift} and LCO filters, and they are given in Table~\ref{tab:specfits}.  Note that the significantly higher spatial resolution of {\it HST} compared to {\it Swift} and seeing-limited images from the ground mean that {\it Swift} and ground-based images will have a significantly higher galaxy fraction in most filters than here.  The model with the power-law continuum has shallower slope than the disc, with a best-fit of $F_\lambda \propto \lambda^{-2.0}$.  This shallower slope reduces the flux needed from the diffuse continuum, and both the optical \ion{Fe}{II} and stellar templates, with the normalization of the last two dropping to zero with this model.  The contribution from the diffuse continuum is strongest around the Balmer and Paschen jumps, in the $U$ and $i$ bands respectively.  We estimate it contributes 13 - 32\% of the flux in the $U$ band and 12 - 30\% of the flux in the $i$ band. Moreover, the flux of the \ion{Fe}{II} complex is a significant fraction of the flux throughout the UV, contributing around 20\% of the flux in the UVW1 and U bands. If this emission reverberates (as has been reported for optical \ion{Fe}{II}, \citealt{barth2013}) then it will also contaminate the accretion disc continuum lags.

\begin{table*}
\centering
\caption{Fraction of flux in each of the spectral components through each filter.  The first numbers in each column are for the power-law model, the second for the disc model.}
\begin{tabular}{cccccc}
Filter & Power-law or disc & Diffuse Continuum & Broad Lines & Fe II & All other components\\
\hline
UVW2 & 0.79, 0.76 & 0.03, 0.08 & 0.11, 0.10 & 0.07, 0.06 & 0.0, 0.0 \\
UVM2 & 0.81, 0.76 & 0.04, 0.11 & 0.025, 0.02 & 0.125, 0.11 & 0.0, 0.0 \\
UVW1 & 0.69, 0.63 & 0.05, 0.14 & 0.05, 0.05 & 0.20, 0.18 & 0.0, 0.0 \\
U (Swift) & 0.64, 0.49 & 0.13, 0.32 & 0.0, 0.0 & 0.22, 0.19 & 0.01, 0.0 \\
B & 0.87, 0.69 & 0.06, 0.16 & 0.06, 0.10 & 0.0, 0.04 & 0.01, 0.01 \\
V & 0.76, 0.54 & 0.08, 0.20 & 0.06, 0.09  & 0.0, 0.07 & 0.10, 0.10 \\
g & 0.77, 0.58 & 0.06, 0.15 & 0.12, 0.18 & 0.0, 0.045 & 0.05, 0.045 \\
r & 0.70, 0.49 & 0.09, 0.24 & 0.15, 0.16 & 0.0, 0.035 & 0.06, 0.075 \\
i & 0.67, 0.42 & 0.12, 0.30 & 0.15, 0.16 & 0.0, 0.01 & 0.06, 0.11 \\
\label{tab:specfits}

\end{tabular}
\end{table*}

\section{Discussion}

We have investigated the variability of the Seyfert galaxy Mrk~110 with an $\approx$ 200 day multi-wavelength campaign. We have measured the lags with respect to the UVW2 band for 10  bands from 0.3-10 keV X-rays with the \textit{Swift} XRT telescope to almost the near-IR. Filtering out the long-term variability ($>10$d), we found that the lags increase with wavelength up to $\approx$ 2.5 days at $\approx$ 9000 \si{\angstrom}. However, below $\approx$ 4000 \si{\angstrom} the lags are very close to zero. On longer timescales we do not have coverage with the \textit{Swift}  XRT but there are variations in the hard X-ray band (15-50 keV) detected by the \textit{Swift}  BAT. Here we find that the g-band lags behind the BAT lightcurve by $\sim10$d and that the other optical bands lag behind the g-band by amounts increasing linearly with wavelength up to almost 10 days for the z-band.

The change of lag behaviour when sampled on different timescales has been noted previously \citep{mchardy2018,cheleouce2019,pahari2020,hernandez} and is most simply explained as emission from more than one reprocessor.

%The presence of multiple physical  components acting  at different timescales is a well known property of accreting objects and it has been thoroughly investigated (mainly with Fourier domain techniques) in the last ten years \citep[e.g.][]{gandhi2010,uttley2011,kara2016,kara2019,mastroserio2019}. Even though the irregular cadence of the data makes Fourier analysis challenging, the wealth of long-term  multi-wavelength AGN campaigns permits to observe similar behaviours  also in these extreme objects \citep{collier1998,mchardy2014,mchardy2018,cheleouce2019,pahari2020,hernandez}.

\subsection{Short timescales}

The increasing continuum lag as a function of wavelength observed on short timescales is consistent with the expectation from an illuminated disc and has been observed also for other accreting super massive black-holes \citep[e.g.][]{wanders1997,collier2001,lira,shappee,edelson2019,cackett2020}. However we highlight here the presence of some interesting features. While for almost all the probed AGN so far, the X-ray to UV lag was always of a few days, \citep[in contradiction with a standard accretion disc predictions; see e.g. :][]{gardner2017,kammoun,Mahmoud2019,cai2018,cai2020}, we  measure a very short and approximately zero lag with respect to the UVW2 band in the range between the X-rays ({UVW2-X-ray lag $\approx$-0.4 -- 1.6 days, 3$\sigma$ confidence}) and near UV ({UVW2-UVW1 lag $\approx$-0.5 -- 0.7 days, 3$\sigma$ confidence}). {Even though the measurements are consistent with simulated standard accretion lags \citep[as also  found recently by ][]{kammoun2021b,kammoun2021}}, we notice that the slope from our best fit is marginally  steeper than expected; such a behaviour might therefore suggest  the presence of a different radial temperature profile induced by a different accretion flow geometry \citep[,][]{wang1999,collier2001,cackett2020}. This, however, does not seem to be confirmed by the rms spectrum, which shows, during the rise, a power-law slope of $\approx$ -4/3 during the first part of the campaign. 

Moreover, despite a clear feature  observed in the {\it HST} spectrum,  we did not find any significant excess in the U band lag ({ UVW2-U lag $\approx$-0.6 -- 1.2 days, 3$\sigma$ confidence}), as often observed in other sources. This excess has been usually associated to the effect of the diffuse continuum from the BLR \citep[see e.g.][]{cackett2018,lawther2018,korista2019}.  

These two elements suggest a fundamental difference between Mrk~110 and the  AGN monitored so far. A first possibility could be the higher accretion-rate of this source, which can affect the geometry of the accretion flow. Recent observations of other high accretion-rate objects \citep[see e.g. ][ for NGC~4769 and Mrk~142 respecitively]{pahari2020,cackett2020}, however, still showed evidence of an X-ray-UV and U band excess: therefore, this parameter alone cannot fully explain the observed discrepancies.  Another possibility could be the presence of a different component in the UV range, affecting the lag. For instance, the  \textit{HST} spectrum shows a clear strong contribution of the DC and the Fe$_{II}$ in the UV, which, however, it is not expected to vary on short timescales. NGC~7469 and Mrk~142 have a smaller mass with respect to Mrk~110 \citep[see e.g.][]{peterson2004,bentzlick2010,bentz2015}. Given that the duration of the campaign for these objects is similar, the diffuse continuum  (DC) from the BLR could have a stronger contribution to the lags from smaller mass objects, leading to a stronger excess in the UV on short timescales. This is also supported by the fact that the lags associated with the emission lines from the BLR in NGC~7469 and Mrk~142 are significantly smaller than the ones measured in Mrk~110  \citep[see e,][]{kollatschny2001,peterson2004,peterson2014,du2014}. On the other hand, we also notice that the short emission lines observed in Mrk~142 are thought to be due to the shadowing of  the BLR by the inner part of the slim disc \citep[][]{cackett2020}. The lower $\dot{m}$ of Mrk~110 (hence its more standard disc configuration), may therefore also explain why the lags are longer. A more detailed and systematic comparison between the different sources is therefore necessary to understand how important the  relative size of the BLR with respect to the disc is for multiwavelength variability studies. %Nevertheless, the variable emission and the lag spectrum are both fully consistent with the predictions of a standard accretion disc. % However, we also  notice that in the UVOT wavelength range a similar lag vs wavelength was found in NGC~4593 \citep{edelson2019}. Moreover, also for this source, the X-ray to UV lag got consistent with 0 when filtering the long-term variations \citet{mchardy2018}. These two sources have in common the presence of strong excess at wavelength shorter than $\approx$ 4000 \si{\angstrom} \citep{landt2011}, suggesting a partial contribution from the BLR continuum also to the lag. %On the other hand, the X-ray to UV lag if the excess is due only to the BLR, should be significantly larger.

 %Mrk~110 is known to show strong  H ($\alpha$,$\beta$,$\gamma$) and He (I and II) lines \citep[see e.g.][]{kollatschny2001,landt2011}. Moreover there is also growing evidence that BLR diffuse continuum could play and significant contribution to the emission and the lags of these objects \citep{korista2001,korista2019,cheleouce2019,netzer2020}. We conclude that the deviations from the accretion disc model, seen in some photometric bands, are due to the presence of an emission line from the BLR.  Evidence of a similar effect was already observed not only with spectroscopically resolved observations of NGC~7469 \citep{collier1998,collier1999} and NGC~5548 \citep{cackett2018} , but also through timescales filtering with continuum reverberation lags for NGC~4593 \citep{mchardy2018}. 
\subsection{Long timescales}
From the analysis of  the long-term variations we observed for the first time a very long lag of $\approx$ 10 days between the hard X-rays and the g band. The response of the lag seems also to get longer, following the same linear relation, as a function of wavelength. Given the large amount of energy from the hard X-ray flare and the very similar shape between the different bands it is reasonable to believe that the long-term optical variations  are indeed  powered by the X-rays. However, despite a larger lag is expected on longer timescales due to the tail in the disc response \citep{cackett2007}, such long X-ray/optical lags are difficult to explain just in terms of a standard accretion disc. This is also supported by the much steeper slope of the rms spectrum ($\lambda F_\lambda \approx \lambda^{-2}$) detected using the smoothed long-term variations. This indicates the presence of a different physical mechanism for the long-term trend. 

Such a long lag could be due to the diffuse continuum from the BLR. For instance,  Mrk~110 is known to have, due to its high luminosity, a large BLR \citep[H$\beta$ emission line lag behind the continuum of $\approx$ 20 light days][]{kollatschny2001,peterson2004}. This would also be supported by the longer lag as a function of wavelength and by the results of our fit, which shows that the stronger contribution of the DC at longer wavelength. However, other two elements points against a scenario where  only the BLR dominates the long term variation: first, during the rise of the long term trend, the flux-flux analysis are fully consistent with the prediction of the accretion disc; second, the HST spectrum shows that the DC can contribute only $\approx$ 10-20\% of the total emission and variability in the optical (Table 5 and Fig. \ref{fig:uvspec}, right panel).

A hint for the solution to such a puzzling behaviour could come from the BAT data. Even though its trend may be affected by the instrument sensitivity, there is a clear difference before and after the peak observed in the hard X-rays. From the lightcurves and the results of the flux-vs-flux analysis it is clear that in the first part of the campaign, while the hard X-rays show a very low flux, the optical and UV band follow the same trend; after the rise in the BAT data, instead, the optical bands show a different response as a function of wavelength, leading to the observed linear trend in the lag spectrum.

\begin{figure}
\centering
\begin{center}

\includegraphics[width=\columnwidth]{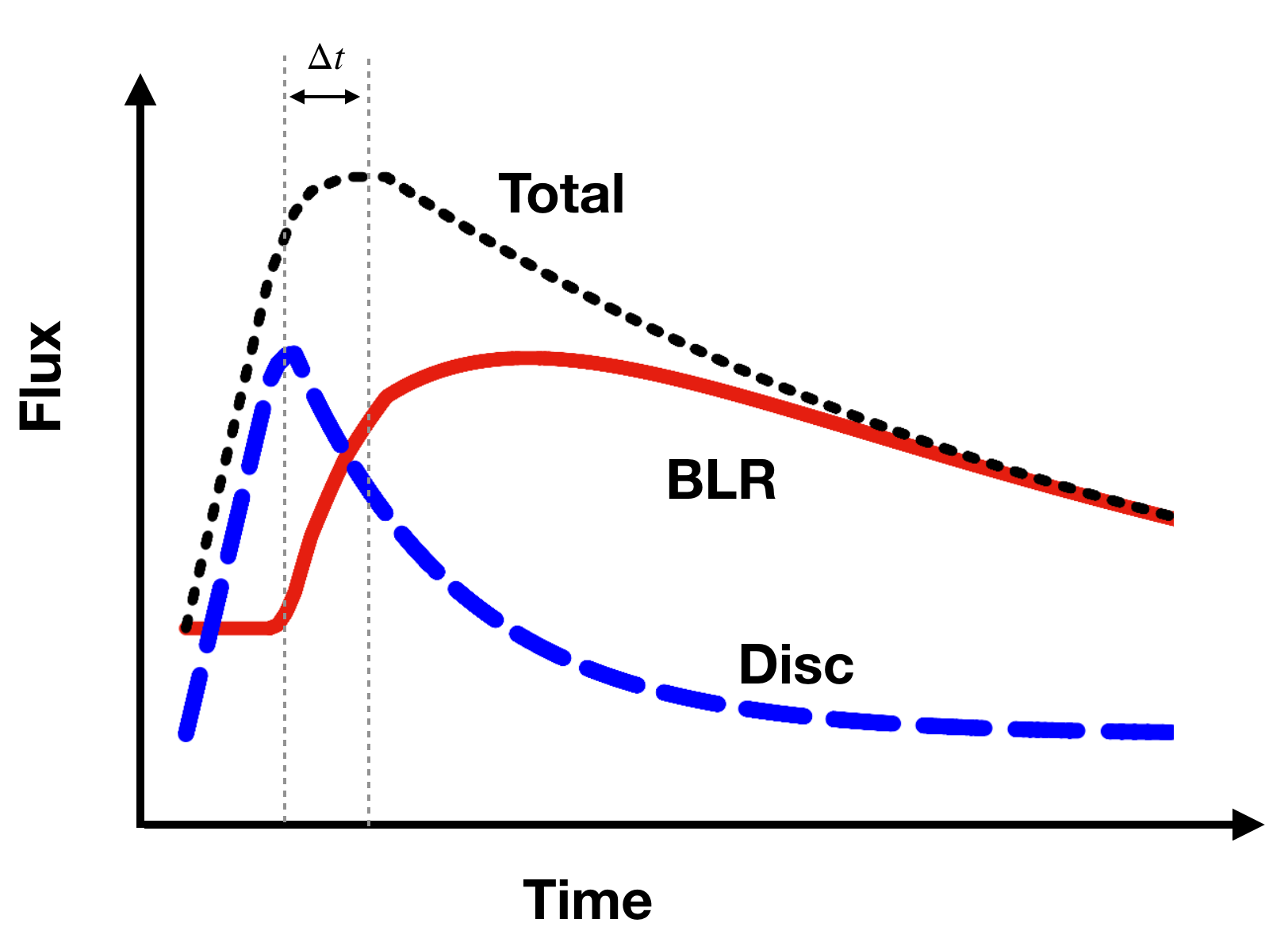}
\end{center}
\caption{Diagram of the proposed physical scenario for the observed long term trend in the optical bands (black fined-dashed curve), assuming a linear rise and a exponential decay. If we assume that the BAT flare is driven by the disc emission (blue long-dashed curve), then this  has a similar rise, an earlier peak (peaks separated by $\Delta$t) , and a faster decay than the optical bands. Therefore in order to recover the total lightcurve, an additional component is needed. We  interpret the difference between the total and the disc lightcurve as the contribution from the DC (red curves), which dominates has the emission after the peak.}
\label{fig:sketch}

\end{figure}

Large variations in the X-rays are usually associated with changes in the geometry of the system \citep{done2007,noda2018}. Spectral timing studies have already shown evidence of the corona changing height on relatively short timescales in both stellar mass and supermassive black holes \citep[see e.g. ][]{kara2019,alston2020,caballerogarcia2020}. A different size or height of the corona would lead to significant changes in the illuminated regions of the accretion flow. This suggests that the observed behaviour is the result of an interplay between accretion disc reflection and of an second reprocessor which can be the DC from the BLR and/or outer regions of the accretion disc. A detailed modelling of the response of the various components is beyond the aim of this paper, however, it is easy to imagine a scenario in which the relative contribution of this reprocessor to the total emission changes according the geometry of the corona (see Fig. \ref{fig:sketch}).

In particular, it is interesting to notice that the DC from the BLR is expected to have a higher contribution during the decay of the X-ray flare, rather than during the rise, not only because of the light travel time distance from the central source, but also because of its size.  Moreover, the response time of the BLR is known to scale as a function of wavelength due to its ionization stratification \citep[see e.g.][]{korista2001,korista2019,lawther2018}.  Therefore this scenario would explain both the slower evolution of the lightcurve at longer wavelengths, as well as the linear trend observed in the lag spectrum (see Fig. \ref{fig:sketch}).

An intriguing alternative possibility could be that the change in measured lag is just a consequence of a change in the temperature profile of the disc. Fig. \ref{fig:lc} shows clearly that the detection of the lag is due to a slower decline at longer wavelength, while the rise is approximately similar at all bands. A variation lasting a few months,  is roughly consistent with the typical disc thermal timescale of these systems: i.e. the timescale on which the disc gets back to thermal equilibrium. It is possible to think that after the BAT flare, the temperature profile of the disc might have steepened its radial profile, leading to a variable spectral slope of -2. For instance, \citet{mummery2019} found that that an accretion disc extending down to the innermost last stable circular orbit with a finite stress, have a spectrum  of $\lambda F_\lambda \propto \lambda^{-12/7} \approx \lambda^{-2}$, so  roughly consistent with what we measured. The main consequence of this would be a clear change in the lag spectrum. However, due to the lack of intensive observations during the tail of the flare, this hypothesis  cannot be  confirmed yet.

New intensive monitoring  observations of this source will help to unveil the nature of this behaviour. In particular these results  show the importance of high cadence spectroscopically resolved observations, which is the only way to disentangle  the different components present in these systems. Future time resolved and wavelength dependent simulations together with further multiwavelength observations can help to understand the origin of these peculiar behaviours.

\section{Conclusions}

We have measured the lag spectrum of the NLS1 galaxy Mrk~110 over a campaign of 200 days. Filtering the data on long ($>10$d) and short ($<10$d) timescales, we find evidence of different behaviour. On short timescales the source shows very short lags which are nonetheless consistent with the expectations of reprocessing from an accretion disc. On long timescales, where the Swift BAT provides the X-ray observations, the g-band lags the hard X-rays by $\approx$ 10 days, with the other optical bands lagging behind the g-band by amounts increasing linearly with wavelength up to almost 10d for the z-band. The simplest, although not the only possible, explanation of this behaviour is direct continuum radiation resulting from reprocessing in the more distant BLR \cite{korista2019}. This behaviour is similar to that seen in NGC~4593 \citep{mchardy2018}. The implied distance to the BLR is longer here than in NGC~4593, but the observed lags are consistent with the $\sim20$d lag of the H$\beta$ line with respect to the continuum \citep{kollatschny2001,peterson2004}. Futher multiband monitoring over long timescales is necessary to properly reveal the contribution of components other than the accretion disc to the UV and optical variability of AGN.

%This seems to indicate the presence of DC from the BLR, as also suggested by spectral measurement. The presence of different processes taking place of at different timescales,  shows therefore the importance of long and continuous monitoring of these sources, especially due to the  non-stationarity of their variability.  Upcoming observations of this same source will allow in the future to shed light on the nature of this intriguing behaviour.

\section*{Data Availability}
\textit{Swift} raw data can be downloaded from NASA heasarc arhive\footnote{\url{https://heasarc.gsfc.nasa.gov/cgi-bin/W3Browse/w3browse.pl}}. \textit{HST} raw data can be downloaded MAST webiste \footnote{\url{https://archive.stsci.edu/hst/}}. The rest of the ground based observations cannot be accessed via web, but are available on request to the corresponding author.

\section*{Acknowledgements}

{The authors thank the referee for the useful comments which improved the quality of paper.}
F.M.V. and I.M.H. acknowledge support from STFC under grant ST/R000638/1.
KH and JVHS acknowledge support from STFC grant ST/R000824/1.
The Liverpool Telescope is operated on the island of La Palma by Liverpool John Moores University in the Spanish Observatorio del Roque de los Muchachos of the Instituto de Astrofisica de Canarias with financial support from the UK Science and Technology Facilities Council. This work makes use of observations from the Las Cumbres Observatory global telescope network. Research at UC Irvine was supported by NSF grants AST-1412693 and AST-1907290.   E.M.C. and J.A.M. acknowledge support for analysis of Zowada Observatory data from the NSF through grant AST-1909199.  E.M.C. and J.A.M. acknowledge support for {\it HST} program number 15413, which was provided by NASA through a grant from the Space Telescope Science Institute, which is operated by the Association of Universities for Research in Astronomy, Incorporated, under NASA contract NAS5-26555. This work makes use of observations from the Las Cumbres Observatory global telescope network.

\bibliographystyle{mnras}
\bibliography{bib_agn2.bib} % if your bibtex file is called example.bib
 
\bsp	% typesetting comment
\label{lastpage}
\end{document}